\documentclass{article}
\usepackage{epsfig,amsmath,amsfonts}

\newcommand{\be}{\begin{equation}}
\newcommand{\ee}{\end{equation}}

\newcommand{\doublet}[2]{\left(\begin{array}{c}#1\\#2\end{array}\right)}
\newcommand{\twobytwo}[4]{\left(\begin{array}{cc} #1&#2\\#3&#4\end{array}\right)}
\newcommand{\threebythree}[9]{\left(\begin{array}{ccc} #1&#2&#3\\#4&#5&#6\\#7&#8&#9\\\end{array}\right)}

\newcommand{\ra}{\rightarrow}

\newcommand{\Rhat}{\hat{\mathrm{R}}}

\newcommand{\R}[4]{R^{#1\phantom{#2}#3}_{\phantom{#1}#2\phantom{#3}#4}}
\newcommand{\Rh}[4]{\hat{R}^{#1\phantom{#2}#3}_{\phantom{#1}#2\phantom{#3}#4}}
\newcommand{\Rt}[4]{\widetilde{R}^{#1\phantom{#2}#3}_{\phantom{#1}#2\phantom{#3}#4}}
\newcommand{\Urm}{\mathrm{U}}
\newcommand{\SU}{\mathrm{SU}}

\newcommand{\SL}{\mathrm{SL}}
\newcommand{\GL}{\mathrm{GL}}
\newcommand{\tq}[2]{\mathrm{t}^{#1}_{\phantom{#1}\!#2}}
\newcommand{\sq}[2]{\mathrm{s}^{#1}_{\phantom{#1}\!#2}}
\newcommand{\Tr}{\mathrm{Tr}}
\newcommand{\Wcal}{\mathcal{W}}
\newcommand{\mparagraph}[1]{\paragraph{#1}\mbox{}\vspace{.1cm}}
\newcommand{\Phibar}{\overline{\Phi}}
\newcommand{\Zset}{\mathbb{Z}}
\newcommand{\Ncal}{\mathcal{N}}
\newcommand{\Fcal}{\mathcal{F}}
\newcommand{\Hcal}{\mathcal{H}}
\newcommand{\Acal}{\mathcal{A}}

\newcommand{\Rcal}{\mathcal{R}}
\newcommand{\AdS}{\mathrm{AdS}}
\newcommand{\Srm}{\mathrm{S}}
\newcommand{\Cset}{{\,\,{{{^{_{\pmb{\mid}}}}\kern-.47em{\mathrm C}}}}}
\newcommand{\Supertwistor}{\Cset \mathrm{P}^{3|4}}
\newcommand{\ket}[1]{\left|#1\right.\rangle}
\newcommand{\qb}{{\bar{q}}}
\newcommand{\hb}{{\bar{h}}}

\newcommand{\comment}[1]{}

\makeatletter
\@addtoreset{equation}{section}
\makeatother

\begin{document}

\begin{flushright}
 AEI-2008-084
\end{flushright}

\vspace{50pt}

\begin{center}

{\LARGE \bf  Quantum Symmetries and Marginal Deformations }\\
\vspace{43pt}

{\large {\mbox{{\bf Teresia M\aa nsson$^{a}$} and {\bf Konstantinos Zoubos$^{b}$}}}}%

\vspace{.5cm}

$^{a}$Max--Planck Institut f\"ur Gravitationsphysik\\

Albert--Einstein--Institut \\

Am M\"uhlenberg 1, D--14476, Potsdam \\
 Germany\\

{\tt teresia@aei.mpg.de}

\vspace{.5cm}

$^{b}$The Niels Bohr Institute\\
Copenhagen University \\
Blegdamsvej 17, 2100 Copenhagen \O \\
 Denmark\\

{\tt kzoubos@nbi.dk}

\vspace{40pt}

{\Large \bf Abstract}

\end{center}

\vspace{.3cm}

\large

\noindent We study the symmetries of the $\Ncal\!=\!1$ exactly marginal 
deformations of $\Ncal\!=\!4$ Super Yang--Mills theory. For generic values
 of the parameters, these deformations are known to break 
the $\SU(3)$ part of the R--symmetry group down to a discrete subgroup.
 However, a closer look from the 
perspective of quantum groups reveals that the Lagrangian is in fact 
invariant under a certain Hopf algebra which is a non--standard quantum 
deformation of the algebra of functions on $\SU(3)$. Our discussion is 
motivated by the desire to better understand why these
 theories have significant differences from $\Ncal=4$ SYM
regarding the planar integrability (or rather lack thereof) of the spin
 chains encoding their spectrum. However, our construction works at the 
level of the classical Lagrangian, without relying on the 
language of spin chains. Our approach might eventually provide a better
 understanding of the finiteness properties of these theories as well as
help in the construction of their AdS/CFT duals.

\normalsize

\noindent

\vspace{0.5cm}

\setcounter{page}{0}
\thispagestyle{empty}
\newpage

\tableofcontents

\section{Introduction}

 One of the most spectacular recent developments in the study of four--dimensional quantum field theory has 
arguably been the gradual uncovering of integrable structures underlying $\Ncal=4$ supersymmetric Yang--Mills (SYM)
theory in the planar limit. The starting point was the observation by Minahan and Zarembo \cite{MinahanZarembo02} 
that, at one loop, the action of the planar dilatation operator on a sector of gauge invariant operators of the theory, 
those formed by traces of a large number of scalar fields, can be mapped to the action of an integrable Hamiltonian on 
states of a particular spin chain. This implies that the problem of diagonalising the dilatation operator (and
thus finding the one--loop spectrum of anomalous dimensions of the theory) can be rephrased as that of finding
the spectrum of energy eigenstates of an integrable spin chain, a problem for which a host of powerful methods
have been developed and can immediately be applied \cite{Faddeev96}. It was soon shown that integrability persists at
higher loops \cite{BKS03} and applies to all sectors of the theory at one loop \cite{Beisert03,BeisertStaudacher03}. 

This conceptual breakthrough in our understanding of $\Ncal=4$ SYM was followed by a large body of work, 
both in the gauge theory as well as on the dual gravity side (describing the regime of strong gauge theory 
coupling according to the AdS/CFT correspondence \cite{Maldacena97}), where similar integrable structures were 
recognised and studied \cite{Mandaletal02,Benaetal03}. The fact that integrability is also seen at strong 
coupling strongly suggests that it persists 
to all orders in perturbation theory, and, although the all--loop dilatation operator (which would
be equivalent to an infinitely--long range integrable Hamiltonian) is not yet known, this assumption of 
integrability made it possible to conjecture a suitable scattering matrix for the spin chain 
degrees of freedom \cite{Staudacher04,Beisert0511} and use it to write down an all--loop Bethe Ansatz 
\cite{BES}, which in principle encodes 
all the information on the planar anomalous dimensions of the theory at any value of the coupling. 
This is a remarkable result, in that
it reduces a complicated field theory problem to the solution of a set of algebraic equations, 
and a large number of checks have helped to refine some of its components (e.g. in checking
a certain scalar dressing factor which cannot be fixed by symmetries) so that it is generally believed that,
for long operators, the complete set of equations is now known. 
 Although there are several outstanding issues which are driving current research, such as that of relaxing the 
constraint of long operators, it is fair to say that the understanding of integrability for $\Ncal=4$ SYM has
reached a high degree of precision and maturity.

However, the theories that the wider scientific community would really like to understand, such as QCD, are quite
far from being integrable. There often exist integrable subsectors at special kinematic
limits\footnote{This was indeed observed in the QCD context much earlier than in the AdS/CFT one \cite{Lipatov93,FaddeevKorchemsky94}.},
but generically the existence of nontrivial dynamics is a clear sign that integrability is 
lost.  Does the discovery of a special four--dimensional gauge theory which is integrable 
have any consequences for these, definitely more interesting, non--integrable field theories?
 
Many researchers in the field would answer affirmatively, the general belief being that $\Ncal\!=\!4$
SYM is a kind of prototype solvable field theory, and that understanding its behaviour in detail 
will provide useful input to the analysis of more realistic, and complicated, quantum field theories. But 
how exactly will this occur?  In trying to understand to what extent the recent advances in our knowledge for
$\Ncal=4$ can teach us something about generic quantum field theories, it is natural to first consider theories 
which are as close  as possible to the $\Ncal=4$ integrable point, without themselves being integrable.

In this work, we will argue that a suitable setting for beginning the study of this question is that
of the exactly marginal deformations of $\Ncal\!=\!4$ SYM, otherwise known as Leigh--Strassler theories.
These theories, which will be reviewed below, arise from $\Ncal\!=\!4$ SYM via superpotential deformations
which break the supersymmetry down to $\Ncal\!=\!1$. Despite the reduced supersymmetry, they 
are believed to be finite to all orders in perturbation theory, in much the same way as $\Ncal\!=\!4$ SYM is (this
will be quantified later). The finiteness property clearly distinguishes these theories as belonging to a very special 
subclass  of four--dimensional field theories, and naturally leads to ask whether there is a special, non--apparent
symmetry which guarantees this finiteness. Supersymmetry is clearly not enough since the generic $\Ncal\!=\!1$ theory
is certainly not finite. 

The story becomes more interesting once we consider the integrability properties of the Leigh--Strassler
theories. As will be discussed more fully in the following, and in marked contrast to $\Ncal=4$ SYM, which is 
believed to be planar integrable in all subsectors (defined by the classes of gauge invariant operators one
considers), its marginal deformations are generically not integrable, though there are special choices of the 
perturbation parameters where integrability is found in certain subsectors. Therefore, having argued that the 
finiteness of the LS theories is possibly a result of a certain hidden symmetry, it seems that this symmetry is 
not strong enough to imply integrability as well.
Note that integrability, at least as it is understood at present, has finiteness (or at least conformality) 
as a prerequisite, because only in conformal theories is the dilatation operator a member of the symmetry
algebra of the theory and can thus be used in classifying states according to their eigenvalues, and eventually
be mapped to a Hamiltonian of a spin chain. But the opposite direction is clearly not true: As evidenced by
the Leigh--Strassler theories, integrability is a much more stringent constraint than finiteness. 

Our proposal for understanding this issue is simply to take a closer look at the symmetries of the theory. In previous
work it has been observed that the Leigh--Strassler theories are closely related to some kind of quantum 
deformation, in the sense of quantum groups, of the $\SU(4)$ R--symmetry group of $\Ncal=4$ SYM. However, precisely 
\emph{which} quantum group, if any, corresponds to the most general Leigh--Strassler theory was never fully 
clarified. By mapping this question to that of characterising the symmetries of a suitable \emph{quantum plane},
we will exhibit a certain Hopf algebraic structure for the general Leigh--Strassler deformation. This can be
thought of as a non--standard deformation of the $\SU(3)$ part of the R--symmetry group. For the special cases
where the gauge theory is known to be integrable, such as $\Ncal=4$ SYM itself and a certain subclass 
of deformations, our algebra becomes dual to a quasi--triangular Hopf algebra. 

Although spin chains are of course an essential part of any discussion of gauge theory integrability, they 
play a slightly secondary role in our approach. Instead of looking at the symmetries of the one--loop spin
chain Hamiltonian, we identify the quantum symmetry directly as the invariance group of the classical 
four--dimensional Lagrangian of the gauge theory.\footnote{Although, as we will
see, the spin chain Hamiltonian will play a significant role in defining the quantum symmetry. More precisely, 
the generators of the algebra will commute with the Hamiltonian and as such the Hopf algebra will directly 
be a symmetry of the spin chain.} Assuming no anomalies at the quantum level, this symmetry
is then naturally a symmetry of the one--loop Hamiltonian as well.

The plan of this paper is as follows: In the following section we review some known features 
of the Leigh--Strassler marginal deformations of $\Ncal=4$ SYM, and in particular what is
known about the integrability properties of these theories. Section \ref{Hopf} aims to 
provide a brief and non--technical introduction to those aspects of Hopf algebras that we
will need in the discussion to follow. After these introductory sections, section \ref{ClassicalHopf}
contains our main assertion, which is that the Leigh--Strassler theories enjoy a certain 
Hopf algebra symmetry which is visible at the level of the classical Lagrangian. In section
\ref{IntegrableCases} we focus on some special cases where the Hopf algebra is enhanced to a dual
\emph{quasi--triangular} Hopf algebra, thus signaling the presence of integrability. In section 
\ref{Noncommutative} we make contact with previous work on how noncommutativity (in the 
sense of star--products) arises in the Leigh--Strassler theories, while section \ref{Conclusions}
contains our conclusions. We have also included an appendix with some fundamental 
Hopf algebra definitions, two appendices containing the derivations of the defining relations 
of our algebra, and
one discussing the possibility of constructing explicit matrix representations of the algebra generators.

\section{Essentials of the Leigh--Strassler deformations} \label{LS}

 In this section, after  briefly reviewing some aspects of the marginal 
deformations of $\Ncal=4$ SYM, both at weak coupling and in the dual strong coupling
description, we discuss what is known regarding integrability for these models. 
 

\subsection{Marginal deformations of $\Ncal=4$ SYM} \label{MarginalDeformations}

 It has long been known that $\Ncal=4$ SYM is contained within a much larger
class of finite four--dimensional quantum field theories, which generically preserve
only $\Ncal=1$ supersymmetry. In $\Ncal=1$ superspace language, these theories can be 
reached by suitable marginal deformations of the superpotential: 
\be \label{LSW}
\Wcal_{\Ncal=4}=g {\mathrm {Tr}}(\Phi^1[\Phi^2,\Phi^3])\longrightarrow \Wcal_{LS}=
\kappa{\mathrm {Tr}}\left(\Phi^1[\Phi^2,\Phi^3]_{q}+ \frac h3\left((\Phi^1)^3+(\Phi^2)^3+(\Phi^3)^3\right)\right)
\ee 
where $[A,B]_q=AB-qBA$ is the $q$--deformed commutator. The gauge group here and in the following 
is taken to be $\SU(N)$. Deforming the theory in this way clearly breaks
the $\Ncal\!=\!4$ supersymmetry to $\Ncal\!=\!1$, as can be seen by considering the R--symmetry: $\Wcal_{N=4}$
is invariant under $\SU(3)\times\Urm(1)_R$ (the largest subgroup of the $\SU(4)$ R--symmetry of $\Ncal\!=\!4$
which is explicit in the $\Ncal\!=\!1$ notation we are using), but turning on both $q$ and $h$ generically 
breaks the $\SU(3)$ component to a discrete subgroup. Note that the space of \emph{classically} marginal 
$\Ncal=1$ deformations is much larger, and can be parametrised by a symmetric three--index tensor $h_{IJK}$ in 
the {\bf 10} of $\SU(3)$, with the deformation of the superpotential taking the form 
$\Tr~h_{IJK}\Phi^I\Phi^J\Phi^K$. However, only a two--parameter subgroup, parametrised by $q$ and $h$, extends to 
quantum finiteness, giving an \emph{exactly} marginal theory.

The finiteness of the marginally deformed theories was first demonstrated order--by--order in perturbation
theory \cite{ParkesWest84,JonesMezincescu84,ParkesWest85,Grisaruetal85,Jones86}, with an all--orders
proof given by Leigh and Strassler \cite{LeighStrassler95} using the NSVZ beta--function
\cite{NSVZ83}.\footnote{There has recently been some controversy regarding the higher--loop finiteness 
of the marginal deformations 
beyond the case of real $\beta$ \cite{Rossietal05,Elmettietal06,Rossietal06,Elmettietal07,Borketal07}. Although in this 
work we do take the traditional point of view, based on the validity of the NSVZ beta--function, at present we are not 
interested in going beyond one loop so our results are not affected by this issue. 
However, it could be that a better understanding of the 
symmetries (which is the main aim of this work) will eventually provide some input into this discussion.} 
 In particular, Leigh and Strassler showed that the condition for finiteness can be parametrised
by a single function of the four couplings $f(g,\kappa,q,h)=0$, where $g$ is the gauge coupling (we set the 
theta angle to zero).  
 One crucial aspect of their proof is that the $(q,h)$--theories enjoy a $\Zset_3$ symmetry cyclically 
permuting the  three chiral superfields. This guarantees that their anomalous dimensions are equal, reducing the number
of variables and guaranteeing that there is a solution for the simultaneous vanishing of all beta functions
and anomalous dimensions. It is precisely this symmetry which singles out the two parameters $q$ and $h$
out of the 10--dimensional space of classically marginal deformations. 

 The function $f(g,\kappa,q,h)$ is not known in general, 
but at one--loop order, and with the above conventions, it reduces to the condition:\footnote{There are 
various derivations and discussions of this condition in the recent literature, e.g. 
\cite{AharonyRazamat02,Razamat02,FreedmanGursoy05,Penatietal05,Borketal07}.} 
\be \label{finiteness}
2g^2=\kappa\bar{\kappa}\left[\frac{2}{N^2}(1+q)(1+\bar{q})+\left(1-\frac{4}{N^2}\right)
\left(1+q\bar{q}+h\bar{h}\right)\right]\;.
\ee
This one--loop condition is actually sufficient to guarantee finiteness at two loop order, and in the
\emph{planar} limit to \emph{three}--loop order (see \cite{Borketal07} for a recent discussion). 
However, even in the planar limit,
at higher loops the requirement of finiteness will eventually modify (\ref{finiteness}). An exception to this 
is the so--called \emph{real $\beta$} deformation, corresponding to $\bar{q}=1/q, h=0$, i.e. $q=\mathrm{exp}(i\beta)$ with
$\beta$ real. It has been shown \cite{Maurietal05} that for this case the resulting
finiteness condition ($\kappa\bar{\kappa}=g^2$) is exact to all orders in planar perturbation theory. 
Since this condition does not depend on $q$, it is the same for the $\Ncal=4$ fixed line at $q=1, h=0$. 
Note however that for $\Ncal=4$ SYM this condition is unmodified when passing to the non--planar level,  
but it will receive non--planar corrections in the real $\beta$ case.

Staying at planar level, we can now be more precise regarding the finiteness properties of 
the Leigh--Strassler theories compared to $\Ncal=4$ SYM: In all cases apart from the real 
$\beta$ deformation, and certain other very special cases which, as we will see, are related to 
it by Hopf twists, the planar one--loop finiteness condition receives corrections at higher loops. 
As has been observed in the past \cite{Borketal07}, and will be discussed further in section 
\ref{IntegrableCases}, exactness of the one--loop finiteness condition is correlated to integrability.

The full $(q,h)$--deformed theory has a left-over $\Zset_3\times\Zset_3$ symmetry (apart from the 
$\Urm(1)_R$ symmetry which is of course preserved by the---$\Ncal=1$ supersymmetric---deformation). 
These $\Zset_3$'s can be taken to act as
\be \label{Z31}
\Zset_3^A:\; \Phi^1\rightarrow \Phi^2\quad,\quad \Phi^2\rightarrow\Phi^3\quad,\quad \Phi^3\rightarrow \Phi^1
\ee
(this is the cyclic permutation symmetry discussed above) and 
\be \label{Z32}
\Zset_3^B:\; \Phi^1\rightarrow \Phi^1\quad,\quad \Phi^2\rightarrow\omega\Phi^2\quad,\quad \Phi^3\rightarrow \omega^2\Phi^3
\ee
with $\omega$ a third root of unity. The two $\Zset_3$'s do not commute with each other and, together with 
another $\Zset_3$ contained within $\Urm(1)_R$ (acting simply as $\Phi^i\ra \omega \Phi^i$), combine to form a 
trihedral group \cite{Aharonyetal02} which will be discussed in more detail in section \ref{Discrete}. In 
\cite{MadhuGovindarajan07b} it was checked that this 
discrete symmetry was preserved in the quantum theory at least up to two loops. We should also note that the 
real $\beta$ deformation preserves a larger  $\Urm(1)^3$  subgroup of $\SU(3)\times\Urm(1)_R$.

Following the influential work of Witten \cite{Witten0312} on the description  of tree--level amplitudes in $\Ncal=4$ SYM 
through a suitable twistor string (a B--model topological string defined on super--twistor space $\Supertwistor$),
it was shown in \cite{KulaxiziZoubos04} that the marginal deformations can be straightforwardly embedded in the
twistor string framework  by introducing a suitable star product among the fermionic directions of $\Supertwistor$,
thus making them non--anticommutative. That work considered only tree--level MHV amplitudes and was restricted to first order
in the deformation parameter. The difficulty in extending to higher orders was linked to the fact that the star product was 
coordinate--dependent, which led to loss of associativity. The approach of \cite{KulaxiziZoubos04} was later 
applied to non--MHV amplitudes in \cite{GaoWu06}. More importantly, for the 
real $\beta$--deformation, the authors of \cite{GaoWu06} were able to show that the star product can be extended 
to all orders in the deformation parameter while preserving associativity.

In the context of amplitudes at loop level, other works discussing perturbative aspects of the $\beta$--deformed 
theories include \cite{Khoze05},
where the equivalence of the gluonic amplitudes to those in $\Ncal=4$ SYM was shown (see also \cite{Ozetal07}), 
as well as \cite{Ananthetal06}, where a light--cone approach was used to demonstrate the all--loop finiteness of the $\beta$ 
deformation.

\subsection{The dual gravity picture}

 The AdS/CFT correspondence \cite{Maldacena97} for $\Ncal=4$ SYM is the conjecture that this theory is
precisely equivalent to Type IIB string theory on $\AdS_5\times\Srm^5$ plus RR five--form flux. A useful limit
of this statement, and the one which has received the most attention, is that where the gauge theory is taken
to be planar, and its 't Hooft coupling to be strong. In that case the dual string theory reduces to classical
IIB supergravity on $\AdS_5\times\Srm^5$, with the $\AdS_5$ part of the geometry parametrising the conformal
invariance of the gauge theory. It is a natural question whether the marginal deformations of $\Ncal=4$ admit a 
similar gravity dual. 
Since conformal invariance is maintained, it is clear that the $\AdS$ part of the gravity background should 
remain, though in principle its radius of curvature could become small, thus invalidating a supergravity approach. 
However, for small (but finite) values of the deformation parameters the deformation should be reflected by a
 suitable deformation of the round $S^5$ geometry which is visible within supergravity.
 
 Some early works which approached the problem perturbatively in the deformation parameter are 
\cite{FayyazuddinMukhopadhyay02,Aharonyetal02}. However, progress in this direction has been hampered by the very 
small amount of leftover symmetry in the full deformation. Generically one only expects to have a single $\Urm(1)$ isometry
direction, corresponding to the R--symmetry of the gauge theory, plus some discrete symmetries as discussed above,
and this symmetry is not enough to construct a useful ansatz which would simplify the solution of the supergravity equations.

 However, in the case of the $\beta$--deformation one expects a residual $\Urm(1)^3$ isometry group, and in this
case Lunin and Maldacena \cite{LuninMaldacena05} showed that one can obtain the dual background by making use of
the two non--R--symmetry $\Urm(1)$'s to perform a sequence of T--dualities and phase shifts. This breakthrough
led to a renewed interest in the properties of the marginally deformed geometries. It is now understood how the
LM background fits into the general framework of $\Ncal=1$ Type IIB flux compactifications, in particular in 
relation to generalised geometry \cite{Minasianetal06,HalmagyiTomasiello07,Granaetal08}. 

 The LM solution also spurred certain attempts to find the geometry dual to the most general $(q,h)$ deformation. 
Inspired by the appearance of non--commutativity in the Lunin--Maldacena approach, the works
\cite{Kulaxizi0610,Kulaxizi0612} attempted to obtain  the full background by
considering the mapping of the open string metric (which is the one seen by the gauge theory, and which exhibits 
non--commutativity) to the closed string one (where the coordinates are commutative, but there is a B--field turned on) 
in the spirit of Seiberg and Witten \cite{SeibergWitten9908}.
 Although this approach (which will be expanded on in section \ref{starproduct}) was successful for 
the case of real $\beta$, it quickly ran into difficulties when applied to the 
full $(q,h)$--deformation, and, as on the twistor string side \cite{KulaxiziZoubos04}, the problem could be traced 
to the non--associativity of the star product for the full deformation. Thus, at present, the construction of
the dual background for the full Leigh--Strassler deformed theory remains an open problem.

\subsection{Marginal deformations vs. integrability}

 The Leigh--Strassler theories, being perturbatively finite, are clearly very special among 
four--dimensional quantum field theories. One can therefore ask whether 
the remarkable properties of $\Ncal=4$ SYM related to integrability extend to its marginal deformations.
In other words, is the property of finiteness enough to guarantee the presence of integrable
structures in the study of the spectrum of the theory? The answer, as 
we now review, turns out to be negative. 

One of the first works to address the issue of integrability in the Leigh--Strassler theories was 
\cite{Roiban03}, which demonstrated one--loop integrability in the $\SU(2)$ subsector for general $q$ 
(with indications that it extends beyond that). The integrable spin chain Hamiltonian describing 
this sector turned out to be a certain parity-violating extension of the XXZ 
Heisenberg spin-chain. As explained in \cite{Mansson0703}, this Hamiltonian is related to the 
Temperley--Lieb generator (see e.g. \cite{Gomezetal}).  
The parity violation does not affect the quantum group symmetry of XXZ, which 
is known to be $U_q(su(2))$ (for its definition, see appendix A), and in any case for a closed spin chain 
the difference between the XXZ
Heisenberg spin chain and its parity breaking--version vanishes. The conclusion is that the Hamiltonian 
of the $q$--deformation in the $\SU(2)$ sector enjoys $U_q(su(2))$ symmetry, or $\SU_q(2)$ in dual language. 

The approach of \cite{Roiban03} was to start from a scalar field theory Lagrangian engineered to produce
a desired integrable spin chain Hamiltonian as its one--loop dilatation operator. The main issue is then 
whether this Lagrangian can be extended by adding ``flavour--blind'' interactions to form part of a full--fledged 
supersymmetric field theory. Berenstein and Cherkis \cite{BerensteinCherkis04} applied similar ideas to 
examine whether the integrable Hamiltonian corresponding to the $\mathrm{SO}(6)$ XXZ model, which
has $\mathrm{SO}_q(6)$ symmetry, can be obtained from a suitable deformation of the $\Ncal=4$ Lagrangian.  
They found a mismatch between the $SO(6)$ XXZ spin chain and the $q$--deformation, which implied that the 
full scalar sector of the $q$--deformed theory was not integrable. Furthermore, although from the analysis
of \cite{Roiban03} in the $\SU(2)$ sector one might be led to expect that the $q$--deformed holomorphic 
$\SU(3)$ sector would also be integrable, they showed that not to be the case unless $q$ is just a phase. 
Finally, they observed that for $q$ a root of unity the one loop dilatation operator in that sector could 
be related through a global transformation to the non--deformed case.

 For $\beta$ real, one--loop integrability in the full scalar field sector (and actually for the larger
$\SU(2|3)$ sector) was later shown in \cite{BeisertRoiban05}. In this case
integrability is also present on the
dual gravity side \cite{Frolovetal05}. Some aspects of higher--loop 
integrability for general $q$ were discussed in \cite{Mansson0711}.

A complete classification of the allowed form for a spin chain Hamiltonian
with  $\Urm(1)^3$ symmetry which had the potential to
describe a three--scalar--field sector (and which excluded the possibility
for a new class of complex 3--parameter Lunin--Maldacena deformations to be integrable) 
was provided in \cite{Freyhultetal05}.

Just as for $\Ncal=4$ SYM, when investigating the integrability of the $(q,h)$--deformed spin chain it
is convenient to restrict to particular subsectors of the theory. Beyond the $\SU(2)$ sector discussed
above, the most obvious such sector is the holomorphic $\SU(3)$ sector, 
where one restricts to single--trace operators composed of the three holomorphic scalars of 
the theory, $\Phi^1,\Phi^2$ and $\Phi^3$. In this holomorphic sector, the
one--loop spin chain Hamiltonian for the general deformation was written down in 
\cite{BundzikMansson05}:\footnote{Due to a different convention, the Hamiltonian appearing in 
\cite{BundzikMansson05} is the transpose of the one here. Our conventions here are such that the Hamiltonian
agrees with the full Hamiltonian written down in \cite{Mansson0703} when restricted to the holomorphic sector.}. 

\be \label{HolomorphicHamiltonian}
H_{l,l+1}=\frac1{1+q\bar{q}+h\bar{h}}\begin{pmatrix}
  h\,\bar{h}&0&0&0&0&\bar{h}&0&-\bar{h}\,q&0\cr 0&1&0&-q&0&0
  &0&0&h\cr 0&0&q\,\bar{q}&0&-h\,\bar{q}&0&-\bar{q}&0&0\cr 0&-\bar{q}&0
  &q\,\bar{q}&0&0&0&0&-h\,\bar{q}\cr 0&0&-\bar{h}\,q&0&h\,\bar{h}&0&
  \bar{h}&0&0\cr h&0&0&0&0&1&0&-q&0\cr 0&0&-q&0&h&0&1&0&0\cr -h\,
  \bar{q}&0&0&0&0&-\bar{q}&0&q\,\bar{q}&0\cr 0&\bar{h}&0&-\bar{h}\,q&0
  &0&0&0&h\,\bar{h}\cr 
\end{pmatrix}\;.
\ee

Associating $(\Phi^1,\Phi^2,\Phi^3)\ra(\ket{1},\ket{2},\ket{3})$, this nearest--neighbour
Hamiltonian is taken to act on the basis given by $\{\ket{11},\ket{12},\ket{13},\ket{21},\ket{22},\ket{23},\ket{31},\ket{32},\ket{33}\}$. 

The Hamiltonian (\ref{HolomorphicHamiltonian}) is not integrable for general values of $(q,h)$. 
The cases where it reduces to an integrable Hamiltonian have been investigated in \cite{BundzikMansson05}. 
Apart from the real $\beta$ case ($\qb=1/q,h=\hb=0$), already discussed in \cite{BerensteinCherkis04,BeisertRoiban05},
and certain roots of unity \cite{BerensteinCherkis04},  it was found that there exist a number of other
integrable cases, but that most of them could be related  by similarity transformations to the 
real $\beta$ case. 

Going beyond the holomorphic sector, integrability for the full scalar field sector was 
investigated in \cite{Mansson0703} using Reshetikhin's integrability criteria, which guarantee 
the existence of an infinite number of commuting charges. Integrability of the holomorphic 
sector was preserved in most cases when extending to the 
full scalar field sector. In that work it was also found that, for
$h=0$ and any complex $q$, there exists an 
integrable subsector with  $U_q(su(3))$ symmetry consisting of 
two holomorphic and one anti--holomorphic scalar (and vice versa), where the 
anti--holomorphic scalar is not conjugate to either of the holomorphic ones.

Having a spectral--parameter dependent $R$--matrix, an integrable spin chain can be recovered via 
a standard procedure (e.g. \cite{Faddeev96}), 
\be
H=-i P \frac{d}{du} R(u)|_{u=0}
\ee
where $u=0$ is the value of the spectral parameter where the $R$--matrix reduces to the permutation operator $P$.  
For all the integrable cases of the holomorphic Hamiltonian 
(\ref{HolomorphicHamiltonian}) there exist rational $R$--matrices, while
the $R$--matrix describing the integrable subsector with two holomorphic
and one anti-holomorphic scalar is trigonometric.

Yangians play a major role for rational integrable models. They are  infinite--dimensional 
Hopf algebras, which provide $R$--matrices with spectral--parameter dependence for the rational models.
 Their appearance in the $\Ncal=4$ context, both at weak and strong coupling, was first 
discussed in \cite{Dolanetal03,Dolanetal04}, with more recent studies focusing on their
role as symmetries of the AdS/CFT S--matrix \cite{Beisert0704,Matsumotoetal07,SpillTorrielli08,Torrielli08}. 
For the real $\beta$-deformed case the Yangian symmetry was discussed in \cite{Ihry08}. 
For \emph{trigonometric} integrable models the corresponding infinite--dimensional
symmetry is an affine quantum group. For example, for the quasi--triangular
Hopf algebra $U_q(su(3))$, introducing an extra parameter to the algebra it is
possible to extend it to an affine quantum group \cite{Gomezetal,ChariPressley}.
 We will only discuss Hopf algebras related
to $R$--matrices without spectral--parameter dependence. However, for the integrable cases we will make 
some connections to the known $R$--matrices with spectral--parameter dependence. It will be interesting
to uncover  a connection between the Hopf algebra we find for
the general case and an affine version or an elliptic quantum group.

In the next section we will introduce Hopf algebras of the type we will later (in
section four) see appearing in the Leigh-Strassler deformations of $\mathcal{N}=4$ SYM.


\section{Introducing Hopf algebras} \label{Hopf}

The plan of this section is to introduce some basic ingredients about
Hopf algebras which will be essential for the analysis in the next section
where we will show how a Hopf algebra structure appears in our physical system. 
For more reading on these basics we refer to e.g. \cite{Majid,Gomezetal,ChariPressley,Chang95}.

\subsection{Quantum linear algebra}

One of the most concrete ways of thinking about quantum symmetries is 
perhaps in terms of quantum linear algebra. Quantum linear algebra works in
analogy with linear algebra. Thus the quantum vector space consists of quantum vectors 
${\bf x}=(x^i)$ and quantum co-vectors ${\bf u}=(u_i)\,,$ 
 where the elements $x^i$ and $u_i$ take their values in a noncommutative space $V$.
Linear transformations are described by quantum matrices ${\bf t}=\{\tq ij \}$,
which can be thought of as ordinary matrices with the difference that the elements $\tq ij$ 
are now operators instead of numbers.

In quantum vector algebra it is common to specify the commutation 
relations between vector elements, and between co-vector elements,
using a matrix $R$ \cite{FRT90}. 
This is a $\Cset$--valued matrix acting on the noncommutative space $V\otimes V$. 
Using the tensor components of the matrix $R$, the relation can be written 
as
\be
\label{vector}
\lambda x^bx^a=R^{ab}_{\; jl}x^jx^l \;,
\ee
\be
\label{co-vector}
\lambda u_au_b=u_ju_i R^{ji}_{\; ba} \;,
\ee
where $\lambda$ is one of the eigenvalues of the matrix 
$\hat{R}^{ab}_{\; kl} := R^{ba}_{\; kl}$, or without indices 
$\hat{R} := PR$. Here $P$ is the \emph{permutation} matrix, $P^{ij}_{kl}=\delta^j_{\;k}\delta^i_{\;l}$ .
Then a quantum symmetry could be considered to be the
transformation of the quantum vector and quantum co-vector which
preserves the forms
(\ref{vector}) and (\ref{co-vector}). Thus the transformations
under consideration are
\be \label{planetransf}
x^{\prime j}= \tq jl x^l\;, \qquad \mbox{and}\qquad u_j'=u_l (t^{-1})^l_ {j} \;.
\ee
Here we have made the assumption that ${\bf t}$ has an inverse (we will soon introduce
a more precise notion, that of an antipode), otherwise the co--plane should have been defined 
in a different way. As will be clear, this choice 
is natural when one is interested in quantum generalisations of $\GL(n)$.
It can be checked that the transformations (\ref{planetransf}) will preserve the 
forms of (\ref{vector}) and (\ref{co-vector}) if the elements $\tq ij$ satisfy
\be
\label{RTT}
R^{i\;k}_{\;a\;b} t^a_{\;j}t^b_{\;l} = t^k_{\;b}  t^i_{\;a} R^{a\;b}_{\;\;j\;\;l}\;,
\ee
where, in performing the calculations, it is  assumed that the elements  $\tq ij$ commute with the vector
and co-vector elements. Equations (\ref{RTT}) go under the name of FRT \cite{FRT90}, or simply RTT,  relations. 
They give rise to what is known as a right/left $\mathcal{A}(R)$-co-module algebra,
where $\mathcal{A}(R)$ will be a bialgebra with generators $\tq ij$ soon to be defined.

\subsubsection{An example} \label{Example}

But first, let us make all this more concrete with an example. The most 
famous one is Manin's  quantum plane:
\be
0=qxy-yx \;, \qquad  \mbox{where} \qquad x=x^1 \qquad y=x^2
\ee
and the corresponding co--plane
\be
0=vw-qwv\;, \qquad  \mbox{where} \qquad v=u_1 \qquad w=u_2\; .
\ee
This quantum plane arises from (\ref{vector}) when using the $U_q(sl(2))$ $R$--matrix:
\be
\label{R-matrix:suq2}
R=q^{-\frac1{2}}\begin{pmatrix}q&0&0&0 \cr 0& 1&q-q^{-1}&0
 \cr 0&0&1&0\cr 0&0&0&q
\cr \end{pmatrix}\; ,
\ee
where the eigenvalue $\lambda=q^{1/2}$ has been chosen.
We may ask whether there exists a linear transformation
\be
{x'}^i=\tq ij x^j \, ,
\ee
which preserves this quantum plane. This is indeed the case, when the elements $\tq ij$ satisfy
\be \label{Aq}
\begin{split}
& \tq 11 \tq 12=q^{-1}\tq 12 \tq 11\;,\; 
\tq 11 \tq 21=q^{-1} \tq 21\tq 11\;,\;
\tq 12\tq 22 =q^{-1}\tq 22 \tq 12\;,\;
\tq 21\tq 22=q^{-1}\tq 22 \tq 21\;,\; \\
& \tq 12\tq 21=\tq 21\tq 12\;,\;
\tq 11 \tq 22-\tq 22\tq 11=(q^{-1}-q)\tq 12 \tq 21 \; ,
\end{split}
\ee
which we leave as an exercise for the interested reader.
The above relations can be deduced from (\ref{RTT}) using the
$R$--matrix (\ref{R-matrix:suq2}).
The matrix  ${\bf t}=\{\tq ij\}$ has many similarities with the 
matrix representation of a group. In particular, assuming that the 
elements ${\tq ij}'$ commute with the elements $\tq ij$, 
then ${\tq lm}''={\tq li}' \tq im$ also represents a generator of the above 
algebra. 

If we now demand the form invariance of the expression 
\be
\label{ch3:forminvariance}
f(x,y):=xy-q^{-1}yx \;,
\ee
under the quantum symmetry, we need to impose an extra constraint on the generators $\tq ij$. 
Defining the \emph{quantum determinant} as 
$\mathbb{D}:=\tq 11 \tq 22 -q^{-1}\tq 21\tq 12$, it can be shown that it 
is \emph{central}, i.e. it commutes with all the generators $\tq ij$
 and we can therefore make a further quotient $\mathbb{D}=1$ of the algebra,
in addition to the quadratic relations.  This defines out of
 the quantum deformation of $\GL(2)$ a quantum deformation of $\SL(2)$. 
 Doing this we obtain that 
$f(x',y')=f(x,y)$. This will be most relevant when constructing the 
Hopf algebra in the next section.

As will be clear from the definition below, the $\tq ij $ are the generators 
of a quantum matrix bialgebra. The special case considered above 
is not just a bialgebra, but a very special Hopf algebra which is dual to a
quasi--triangular Hopf algebra, the universal enveloping algebra
$U_q(sl(2))$. See appendix A for the basic definitions of bialgebras
and Hopf algebras.

\subsection{Quantum matrix algebra}

We now discuss how the general definitions of bialgebras and Hopf algebras 
in Appendix A apply to the matrix algebra case. In the following $M_n$ is the space of $n\times n$ matrices.

\mparagraph{Quantum matrix bialgebra}

Let $R$ be an element of $M_n\otimes M_n$. The bialgebra $\Acal(R)$
of quantum matrices is defined as being generated by 1 and $n^2$
indeterminates ${\bf t}=\{t^i_{\;j}\}$ with 
\be
\label{relations:1}
R^{i\;k}_{\;a\;b} t^a_{\;j}t^b_{\;l} =  t^k_{\;b}  t^i_{\;a} R^{a\;b}_{\;\;j\;\;l},\qquad 
\Delta t^i_{\;j} =\sum_a t^i_{\;a}\otimes t^a_{\;j}, \qquad \epsilon t^i_{\;j}=\delta^i_{\;j}
\ee
where $\Delta$ is the comultiplication operator and $\epsilon$ the counit (see Appendix A). 
Note that the multiplication in the above example, 
 ${\tq lm}''={\tq li}' \tq im$, where the elements ${\tq ij}'$ were
assumed to commute with the elements ${\tq ij}$, is nothing but a 
realisation of the co-product $\Delta$.

It will be useful to think of the algebra $\mathcal{A}(R)$ as a quotient algebra of a free
algebra,  $\mathcal{A}(R)=\Cset[[\tq ij]]/\mathcal{I}$, where $\mathcal{I}$ is the
ideal generated by the quadratic relations coming from the 
RTT relations (the first relation in (\ref{relations:1})).

\mparagraph{Quantum matrix Hopf algebra}

The above bialgebra becomes a Hopf algebra if there exists, for a given
 matrix ${\bf t}$, a matrix, denoted by ${\bf s}=\{\sq ij\}$, which satisfies
$\sq ik \tq kj =\delta^i_{\;j}=\tq ik\sq kj $. In that case the mapping 
$S(\tq ij)\rightarrow \sq ij$ will satisfy the axioms for an antipode.
For the example above an explicit expression can be written down
for the antipode in terms of linear combinations of the generators 
$\tq ij$, demonstrating that they represent a Hopf algebra. Moreover the 
$R$ in (\ref{R-matrix:suq2}) satisfies the Yang-Baxter equation (without
spectral--parameter dependence)
\be
\label{ch3:YB}
R_{12}R_{13}R_{23}=R_{23}R_{13}R_{12} \quad (\text{in index notation:}
\quad \R isjr \R slkp \R rmpn = \R jskp \R irpn \R rlsm)\, .
\ee
When $R$ satisfies the Yang--Baxter equation (YBE)
one is guaranteed that the algebra is not too trivial and it can be shown 
to be dual to a quasi--triangular Hopf algebra. As is well known (e.g. \cite{Faddeev96,Gomezetal}), this is the case
that corresponds to integrable systems.

As a bialgebra, $\Acal(R)$ is perfectly well defined even without the matrix $R$ satisfying the 
Yang-Baxter equation. But when taking $R$ to be an arbitrary matrix one is
not guaranteed that the first relation in (\ref{relations:1}) has any 
solutions (except for ${\bf t}$ being the identity matrix). This is because of the  
large number of equations which are obtained from this relation: If $n$ is the number of 
generators, the number of equations will be $n^2$, but only $n(n-1)/2$ of those should be
independent commutation relations.
Even if for a given matrix  $R$ there exists a non--trivial solution for the quadratic 
relation, one also obtains extra cubic relations when $R$ does not satisfy
 Yang-Baxter. There are two different ways to obtain cubic relations, which follow 
from applying the quadratic relations in different order: Either
\be
\label{ch3:YB1}
R_{12}R_{13} R_{23}{\bf t}_1{\bf t}_2 {\bf t}_3=R_{12}R_{13}{\bf t}_1{\bf t}_3{\bf t}_2R_{23}
=R_{12}{\bf t}_3{\bf t}_1{\bf t}_2R_{13}R_{23}=
{\bf t}_3{\bf t}_2 {\bf t}_1 R_{12}R_{13} R_{23}
\ee 
or
\be
\label{ch3:YB2}
 R_{23}R_{13}R_{12}{\bf t}_1{\bf t}_2 {\bf t}_3=R_{23}R_{13}{\bf t}_2{\bf t}_1{\bf t}_3R_{12}
=R_{23}{\bf t}_2{\bf t}_3{\bf t}_1R_{13}R_{12}=
{\bf t}_3{\bf t}_2 {\bf t}_1 R_{23}R_{13}R_{12}\, .
\ee 
In deriving (\ref{ch3:YB1}) and (\ref{ch3:YB2}) we used the fact that the algebra is associative
(by definition). We see that the YBE guarantees that both orderings lead to equivalent relations. 
Combining the two equations we find:
\be
(R_{12}R_{13} R_{23}- R_{23}R_{13}R_{12}){\bf t}_1{\bf t}_2 {\bf t}_3=
{\bf t}_3{\bf t}_2 {\bf t}_1 (R_{12}R_{13} R_{23}- R_{23}R_{13}R_{12})\,.
\ee
Clearly this equation is automatically satisfied when the matrix $R$ is 
in the same equivalence class as an $R$--matrix which satisfies the  
Yang-Baxter equation.
If this is not the case the equation leads us to extra cubic relations,
which make the algebra inconsistent as a quadratic algebra (see e.g. 
\cite{deAzcarragaRodenas95}
for a discussion). 
This means that the ideal generated by the quadratic relations also leads to at least 
cubic relations and maybe even higher order relations.
This makes the ideal of the algebra larger, and thus the algebra more trivial. 
In \cite{deAzcarragaRodenas95} there is a simple example of how the
cubic constraints follow from the quadratic ones
by performing the multiplication of a cubic term in two different orderings.

Equivalently to the defining relations (\ref{relations:1}) we could also 
have used the permuted $R$--matrix to define the algebra $\hat{R}=PR$, or,
written in index notation  $\hat{R}^{ij}_{kl}=R^{ji}_{kl}$, as follows 
\be
\label{relations:2}
\hat{R}^{i\;k}_{\;a\;b} t^a_{\;j}t^b_{\;l} = t^k_{\;b}  t^i_{\;a} \hat{R}^{b\;a}_{\;\;j\;\;l}\,.
\ee
Actually, we could always add a matrix proportional to the 
identity matrix to the matrix $\hat{R}$ and it would still give us 
the same algebra, or equivalently we could add a matrix 
proportional to the permutation matrix to the matrix $R$ in 
(\ref{relations:1}). In formulas, we can express this as
\be
\label{ch3:equivalence}
\hat{R}\sim aI+b\hat{R}\, , \qquad \mbox{where}\quad a,b \in \Cset,
b\neq 0 
\ee
where $\sim$ denotes that they belong to the same equivalence class.
This means that there is a large equivalence class of 
$R$--matrices which generate the same algebra. Also, even if an 
$R$--matrix does not fulfil the YBE it will still generate an algebra dual to a 
quasi--triangular Hopf algebra, as long as it belongs to the same class
as one that does.
From equation (\ref{relations:2}) we see that if we could think of $\hat{R}$
as the nearest neighbour interaction of a Hamiltonian, the matrix 
$U=t\otimes t$ looks like a symmetry (here the tensor product refers to a 
normal matrix tensor product and not as in the definition of the bialgebra).
 For instance, as discussed in \cite{Mansson0703}, the spin--chain 
Hamiltonian describing the dilatation operator for the $\SU(2)$ sector
can be written as
\be
\label{ch3:Hamiltonian}
H=\sum_i^L e_i \, ,
\ee
where $e_i$ is the Temperley--Lieb generator acting on spin sites
$i$ and $i+1$, which is related to the $R$--matrix (\ref{R-matrix:suq2}) 
as follows
\be
e_i=qI-q^{1/2}\hat{R} \;.
\ee
Thus the Hamiltonian (\ref{ch3:Hamiltonian}) commutes with ${\bf t}^{\otimes L}$
and the Temperley--Lieb generator $e_i$ is in the same equivalence class as
$\Rhat$. In the same way the full holomorphic spin chain Hamiltonian 
(\ref{HolomorphicHamiltonian})
representing the planar one--loop dilatation operator is related to $\Rhat$,
which describes (as we will show) the Hopf algebra describing the symmetry 
of the Leigh-Strassler deformation. When doing the Hopf algebra calculations
in appendices B and C, we will find it convenient to use $\Rhat$, because
of the simple way it is expressed in terms of the tensor $E_{ijk}$
that will be defined in the next paragraph.

\subsection{The three-dimensional quantum plane}

Let us now  briefly review some aspects of work by Ewen and Ogievetsky
\cite{EwenOgievetsky94}, which has provided the inspiration for much of our approach.
That work is concerned with the classification of three--dimensional quantum 
planes, defined as a polynomial algebra with three elements obeying quadratic 
relations such that the Poincar\'e series of the algebra coincides with 
the classical one.\footnote{This essentially means that the number of relations obeyed by the quantum algebra
generators at every degree (quadratic, cubic, etc.) is the same as in the classical algebra.}
Ewen and Ogievetsky find that for  three--dimensional planes
this condition is equivalent to the  matrix $R$ generating the quantum plane satisfying 
the YBE.
They start out by defining the quadratic relations
\be \label{quadrrels}
E^{\alpha}_{ij}x^ix^j=0, \qquad u_iu_j F^{ij}_{\; \alpha}=0 \;.
\ee
Demanding that the independent relations be the same as in the classical case 
they obtain three linearly independent relations.
They introduce two tensors $E_{ijk}$ and $F^{ijk}$  defining the
 quantum plane and  the quantum co--plane respectively:
\be
\label{ch3:3DQplanes}
E_{ijk}x^ix^jx^k=0\,, \qquad \mbox{and} \qquad u_i u_j u_k F^{ijk}=0 \;.
\ee
These tensors are related to the ones in (\ref{quadrrels}) through:
\be
E_{ijk}=E^{\alpha}_{ij}f_{\alpha k}, \qquad \mbox{and} 
\qquad E_{lij}=e_{l \alpha }E^{\alpha}_{ij}\;,
\ee
where the $f_{\alpha k}$ and $e_{l \alpha }$ are related to the cyclicity
properties of the $E_{ijk}$ tensor as follows
\be
E_{ijk}=Q^l_k E_{lij}\, , \qquad \mbox{with} \quad 
Q^i_j=f_{\alpha j}(e^{-1})^{\alpha i} \,.
\ee
We will be interested in the case when $Q$ is the identity matrix such
that $E_{ijk}$ becomes periodic in the indices. As will be clear, this is forced upon us 
by the physical system we have in mind, and in particular the wish to 
preserve the $\Zset_3$ symmetry of the superpotential mentioned above. For similar reasons
we also want the co--plane to have the same nonzero components as the plane.
The condition provided in \cite{EwenOgievetsky94} for $\hat{R}$ to generate the appropriate
algebra is the following
\be \label{EORhat}
\hat{R}^{ij}_{\; kl}=\delta^i_k \delta^j_l - E_{klm} F^{mij}\;,
\ee
where $E_{ijk}$ and $F^{ijk}$ need to satisfy
\be
\label{ch3:normalisationEF}
\delta^i_j=\frac 12 E_{jmn}F^{mni}\;,
\ee
and
\be
 E_{ajm}F^{mib}E_{ebk} F^{kcj}=
{\delta}^c_a \delta^i_e +\delta^i_a \delta^c_e \;. 
\ee
Apart from \cite{EwenOgievetsky94}, these equations were later studied in detail 
in \cite{Ohn02} (see also \cite{Ohn97} for some background)
in order to classify the $\SL(3)$ cases, and it was found they only
have a solution in exceptional cases. We should point out that we have rescaled the $E_{ijk}$ 
and $F^{ijk}$ tensors relative to \cite{EwenOgievetsky94}. In particular, their formula equivalent to
(\ref{EORhat}) would have a 2 in front of $E_{klm}F^{mij}$. This is just a choice of normalisation of the 
tensors and  has no real significance. On the other hand, once we have fixed this normalisation (e.g. by 
requiring (\ref{ch3:normalisationEF})) the relative factor between the first identity term 
and the second term in (\ref{EORhat}) is important for the $R$--matrix to satisfy the YBE. 
But in the subsequent discussion we will relax the Yang-Baxter condition, because our main 
interest is to find an interesting  Hopf algebra (not necessarily dual 
quasi--triangular). We will also not be concerned with
preserving the classical number of independent relations at each degree, or equivalently the
Poincar\'e series condition of \cite{EwenOgievetsky94}.

The quantum determinant $\mathbb{D}$ for three--dimensional quantum planes is defined through the
tensor $E$ \cite{EwenOgievetsky94}: 
\be
\label{ch3:QD}
E_{ijk} \tq il \tq jm \tq kn =\mathbb{D} E_{lmn} \;.
\ee
This should be read as a condition for the invariance of the quantum plane
as well as a natural definition for the quantum determinant, because if we set $\mathbb{D}=1$ 
(which is only possible if it is central), we obtain the condition for a quantum deformation of
$\SL(3)$ instead of $\GL(3)$, just as in the classical case.  
Another expression for the quantum determinant follows from (\ref{ch3:QD}) by 
contracting it with $F$ from the right:
\be
\label{ch3:QDdef}
\mathbb{D}=\frac1{6}E_{ijk}\tq il \tq jm \tq km F^{lmn}\;.
\ee

\subsection{Hopf algebra twists}

 An important property of quasi--triangular Hopf algebras is that they can be \emph{twisted}
to produce new quasi--triangular Hopf algebras with a twisted R--matrix and corresponding
twisted coproduct. To define twisting, one starts with a \emph{counital 2--cocycle}, an element 
$\Fcal\in \mathcal{H}\otimes \mathcal{H}$ which is invertible and satisfies\footnote{As shown by Drinfel'd \cite{Drinfeld89}, 
these conditions on $\Fcal$, including invertibility, 
can actually be relaxed in a useful way. Twists with these more general $\Fcal$'s would take us out of the 
regime of Hopf algebras to that of (non--associative) quasi--Hopf algebras.} 
\be
\label{ch3:axiomsF1}
(\epsilon\otimes \text{id})\Fcal=1=(\text{id}\otimes \epsilon)\Fcal
\ee
and 
\be
\label{ch3:axiomsF2}
(1\otimes \Fcal)(\text{id}\otimes\Delta)\Fcal=(\Fcal\otimes 1)(\Delta\otimes \text{id})\Fcal \;.
\ee
We are only interested in the effect of the twist on the original R--matrix. It is given 
by
\be
\label{ch3:twist}
R'_{12}=\Fcal_{21}R_{12}\Fcal^{-1}_{12}\;.
\ee
Upon expressing the above equation for $\hat{R}$ we see it takes the form
\be
\hat{R}'_{12}=\Fcal_{12}\hat{R}_{12}\Fcal^{-1}_{12}\,.
\ee
Here we see that $\Fcal$ acts as a similarity transformation on $\hat{R}$, but
recall that it does not mean that the algebras will be isomorphic, because
for that to happen the transformation needs to act separately in the
${\bf t_1}$ and ${\bf t_2}$. A normal change of basis of the generators
$\tq ij$ in the dual algebra would correspond to
a $\Fcal$  written in the form $U\otimes U$ with
the matrices $U$ being the same.

Note that the axioms for $\Fcal$ are consistent with the axioms for $\Rcal$
to define a quasi--triangular Hopf algebra (\ref{AxiomsR}), that is a matrix 
$\Fcal$ that satisfies the axioms (\ref{AxiomsR}) can be shown to also satisfy
the axioms (\ref{ch3:axiomsF1})  and (\ref{ch3:axiomsF2}). Taking two different twists,
one can ask how do we know that they will generate 
two genuinely different Hopf algebras which are not isomorphic? This depends
on whether the twists are in the same cohomology class \cite{Majid}
\be
\Fcal^{\gamma}=(\gamma \otimes \gamma)\Fcal \Delta \gamma \;.
\ee
In section {\ref{IntegrableCases}} we will make use of the twist transformation to relate
the quasi--triangular Hopf algebras associated to several integrable Leigh--Strassler 
deformations.


\section{Hopf Symmetry of the classical Lagrangian} \label{ClassicalHopf}

In the following we will discuss a particular Hopf symmetry which is the invariance
symmetry of the one--loop planar dilatation operator, or equivalently the Hamiltonian 
(\ref{HolomorphicHamiltonian}).  In particular, the generators $\tq ij$ of the 
Hopf algebra represented by the matrix ${\bf t}$ will turn out to satisfy
\be \label{Hinvar}
{\bf s}^{\otimes L} H {\bf t}^{\otimes L}=H \,,
\ee
where $H=\sum_{i=1}^{L} H_{i,i+1}$ and ${\bf s}$ the antipode discussed above.
However, the viewpoint we would like to take in this work is that
this Hopf symmetry of the one--loop Hamiltonian is actually already present at the level of the 
classical Lagrangian. 

 An early indication that there exists a quantum symmetry related to the general Leigh--Strassler
theory appeared in the work of \cite{Berensteinetal00}. Those authors noticed that the 
moduli space of vacua of the theory (obtained by minimising the potential) 
has a (cyclic) quantum plane structure:
\be\label{cyclicplane}
\begin{split} 
\phi^1\phi^2&=q\phi^2\phi^1-h (\phi^3)^2\\
\phi^2\phi^3&=q\phi^3\phi^2-h (\phi^1)^2\\
\phi^3\phi^1&=q\phi^1\phi^3-h (\phi^2)^2
\end{split}
\ee
where $\phi^i$ denotes the expectation value of the scalar part 
of $\Phi^i$. Correspondingly we could write the conjugated relations, 
defining a cyclic co--plane. As discussed earlier, one possible definition
of quantum groups is as the symmetry groups of quantum planes. Thus, by 
considering the geometry of the moduli space we see that there should
be an appropriately defined quantum group acting on it. However, the work of 
\cite{Berensteinetal00} did not specify precisely which quantum symmetry 
corresponds to the general $(q,h)$ deformation. 

Motivated by \cite{Berensteinetal00}, in the following we will explore the
symmetries of the quantum plane in (\ref{cyclicplane}). However, we will be
even more general, and will ask which are the quantum transformations which leave the
\emph{superpotential itself} invariant, rather than just its space of solutions.

\subsection{Deforming the superpotential}

We will start by exhibiting the full deformed superpotential, 
with both $q$ and $h$ nonzero, in a form which will help to make the relation
to Hopf algebras, in the way discussed in the previous section, obvious.
This will result in a two parameter deformation of the $su(3)$ algebra.
 
Let us start from the $\Ncal=4$ superpotential:
\be 
\Wcal_{\Ncal=4}=g\Tr\left\{\Phi^1 [\Phi^2,\Phi^3]\right\}=
\frac g3 \epsilon_{ijk} \Tr\left\{\Phi^i\Phi^j\Phi^k\right\}\;.
\ee
Here the superpotential is expressed via the $\SU(3)$--invariant
tensor $\epsilon_{ijk}$. We would now like to see the Leigh--Strassler
superpotential as arising from deforming the $\epsilon_{ijk}$ tensor to $E_{ijk}$, 
a tensor invariant under a \emph{quantum} deformation of $\SU(3)$. The goal is 
to prove its invariance under some generators ${\bf t}$, which form
a Hopf algebra, as explained in section 3. We will also of course  need 
invariance of  the hermitian conjugate of the superpotential, which will 
define for us the co--tensor $F^{ijk}$.
Let us use the trace structure to write the Leigh-Strassler superpotential (\ref{LSW}) as 
\be \label{LSexpanded}
\Wcal_{LS}=\frac{\kappa}{3} \Tr\left\{\Phi^1 \Phi^2\Phi^3\!\!+\!\Phi^2 \Phi^3\Phi^1\!\!+\!\Phi^3 \Phi^1\Phi^2
\!\!-\!q(\Phi^1 \Phi^3\Phi^2\!\!+\!\Phi^2 \Phi^1\Phi^3\!\!+\!\Phi^3 \Phi^2\Phi^1)
\!+\!h[(\Phi^1)^3\!\!+\!(\Phi^2)^3\!\!+\!(\Phi^3)^3]\right\}.
\ee
Our main interest is the form invariance of the superpotential which is related to 
the three--dimensional quantum plane, in an analogous way to the example in the
previous section exhibiting the relation between Manin's quantum plane and the 
form invariance in (\ref{ch3:forminvariance}).
To investigate form invariance, we express the superpotential and
its hermitian conjugate in terms of the tensors $E_{ijk}$ and $F^{ijk}$ as 
\be
\Wcal_{LS}+{\Wcal}^{\dagger}_{LS}=\frac13 \Tr \left(E_{ijk}\Phi^i\Phi^j\Phi^k+
\Phibar_i\Phibar_j\Phibar_k F^{ijk}\right).
\ee
Comparing with (\ref{LSexpanded}), and using the notation of  \cite{EwenOgievetsky94}, 
we find:
\be
\begin{split} \label{qhchoice}
& F^{ijk}= \overline{E}_{ijk} \qquad \mbox{(the bar denotes complex conjugation)},\quad\\
& E_{123}=E_{231}=E_{312}=\frac1{d}\;, \\
& E_{321}=E_{213}=E_{132}=-\frac{q}{d}\;, \\
& E_{111}=E_{222}=E_{333}=\frac{h}{d}\,,\qquad \text{where} \quad d^2=\frac{1+\bar{q}q+\bar{h}h}{2}\;.
\end{split}
\ee
where the normalisation is such that equation (\ref{ch3:normalisationEF}) in section \ref{Hopf}
is satisfied. However, comparing with the finiteness condition (\ref{finiteness}), we find that
the coefficient in front of the superpotential is precisely what is required by planar finiteness, 
in other words $\kappa=1/d$. Recall that (\ref{ch3:normalisationEF}) was required to obtain an 
$R$--matrix satisfying the YBE, but since we will be working in a more general setting we 
are in principle free to choose the normalisation of the tensors 
$E_{ijk}$ and $F^{ijk}$. It is however a peculiar coincidence that the most natural way to choose 
the normalisation agrees with what is obtained for the planar finiteness condition.
 As we will see, this normalisation also has its advantages when expressing the quantum determinant. 
Note that in our discussion of the algebra below we will not assume that we are in the planar limit. 

These choices for $E$ and $F$  were included in the analysis of \cite{EwenOgievetsky94}, 
even though the condition to fulfil the classical Poincar\'e series was too
strong for generic values of $q$ and $h$ to be included in their definition
of a quantum plane.
The case of arbitrary $q$ and $h=0$ was included for the quantum plane
but with a different co--plane, and similarly for the case of arbitrary $h$ and $q=0$. 
For the case $h=0$, $E_{ijk}$ is proportional to the $q$--epsilon tensor
as defined in Majid \cite{Majid}. 

Let us now recall that the component scalar field
part of the F-term Lagrangian can be written as \cite{Roiban03}:
\be
\mathcal{L}_{F,s}=\Tr \overline{\phi}_i\overline{\phi}_jH^{ij}_{lm}\phi^l \phi^m
\ee
where $H^{ij}_{lm}$ are the components of the hermitian matrix
$H$, given explicitly in (\ref{HolomorphicHamiltonian}), describing the local action of the one-loop 
dilatation operator on nearest neighbours, 
\be
H=H_{mn}^{jk}e_{jm}\otimes e_{kn} \qquad \mbox{where} \qquad
H_{mn}^{jk}=E_{mna}F^{ajk}\,.
\ee
Here we introduced the operators $e_{mn}$, which are  defined through their 
action on the spin state $|k\rangle$, as $e_{mn}|k\rangle =\delta_{nk}|m\rangle$.

We would like to show that there exists a quantum algebra transformation acting on $\Phi^i$ as
\be
\Phi^i \rightarrow \tq ij \Phi^j
\ee
under which the deformed superpotential is invariant. Invariance of the superpotential implies that 
\be
E_{ijk}\Phi^i\Phi^j\Phi^k\rightarrow E_{ijk} \tq il \tq jm \tq kn \Phi^l\Phi^m\Phi^n=E_{lmn}\Phi^l\Phi^m\Phi^n
\ee 
i.e. that the $\tq ij$ generators we are interested in finding satisfy 
\be \label{ch4:invariance1}
E_{ijk} \tq il \tq jm \tq kn=E_{lmn}\;.
\ee
A similar condition arises by requiring invariance of the hermitian conjugate 
of the superpotential:
\be \label{ch4:invariance2}
\Phibar_i\ra \Phibar_j{\tq ji}^*\; \quad \Rightarrow \quad {\tq li}^* {\tq mj}^* {\tq nk}^* F^{ijk}=F^{lmn} \;.
\ee
These relations impose strong restrictions on the generators $\tq ij$ of the algebra,
which, as we will see, are compatible with the cubic relations derived for our
Hopf algebra in Appendix C.
The condition (\ref{ch4:invariance1}) above should be compared to the condition 
(\ref{ch3:QD}) for the three dimensional quantum plane in the previous section, 
from which it follows that the quantum determinant occurring in (\ref{ch3:QD}) should 
be set to one.

Since the non--abelian nature of the scalar superfields is not relevant for the following 
discussion (the generators of $\SU(N)$ being taken to commute with the $\tq ij$) from now on
we will return to the quantum plane notation of section \ref{Hopf} and look at the form invariance 
of the expression 
\be
f(x,y,z):=E_{ijk}x^ix^jx^k \,
\ee
where we have associated each of the three holomorphic scalars to one of the quantum plane coordinates. 

As discussed, setting the quantum determinant to one is just the step passing from a quantum
deformation of $\GL(3)$ to that of $\SL(3)$. However we also need to require form invariance of
\be
g(\bar{x},\bar{y},\bar{z}):=\bar{x}_i\bar{x}_j\bar{x}_k F^{ijk}=
\bar{f}(\bar{x},\bar{y},\bar{z})
\ee
which assures the reality condition. We will see that this results in a deformation of $\SU(3)$ instead of $\SL(3)$.

Form invariance for $f(x,y,z)$ and $g(x,y,z)$ together is equivalent to
invariance of $H^{ij}_{kl}$. $H^{ij}_{kl}$ is an hermitian operator
which gives the reality condition for the Hopf algebra that, as we will show,
is generated by it. It is necessary for the existence of a ${\tq ij}^*$
generator.

A Hopf algebra of this type is called \emph{real type} if the following condition on 
the $R$--matrix is satisfied:
\be
\label{ch4:realitycondition}
\overline{\R ikjl}=\R ljki
\ee
which is equivalent to that $\Rh ijkl=\R kjil$ is hermitian as a $9\times 9$ matrix. Here 
 $H^{ij}_{kl}$ will play the role of  $\Rh ijkl$, and since $H$ is hermitian (\ref{HolomorphicHamiltonian})
we are guaranteed to obtain an $R$--matrix of real type.  When $R$ is of real type
the definition 
\be
\label{ch4:def.antipode}
{\tq ij }^*=S(\tq ij)=\sq ij
\ee
is compatible with the relations $R {\bf t_1 t_2}={\bf t_2 t_1}R$ of the
Hopf algebra $\Acal(R)$. So, as in our example in section \ref{Example},
the co--plane coordinates transform according to the antipode. 

Our first question is now whether we can have a non--trivial bialgebra generated by 
  $H^{ij}_{kl}$ as explained in the previous section, i.e. whether there
exists a a non--trivial solution to
\be
\label{Hsymmetry}
H^{ij}_{kl}\tq km \tq ln =\tq ik \tq jl H^{kl}_{mn} \;.
\ee
Note that the same algebra can equally well be generated by any $\hat{R}$ matrices 
belonging to the same equivalence class as $\hat{R}^{ij}_{\; kl}=H^{ij}_{\; kl}$, 
equation (\ref{ch3:equivalence}). 

If this is the case, we would then like to show the existence of an antipode from which it will 
also follow that the superpotential is invariant, since it will imply that the quantum determinant is central.
At the same time, having an antipode will imply (\ref{Hinvar}) and thus guarantee invariance of the spin
chain Hamiltonian under the Hopf algebra. 
We will turn to the analysis of (\ref{Hsymmetry}) after first exhibiting the $R$--matrix related to 
our Hamiltonian. 

\subsubsection{The $R$--matrix for the general deformation}

For concreteness, let us give here the form of the $R$--matrix that follows from the 
choice (\ref{qhchoice}) via (\ref{EORhat}): 
\small  
\be \label{qhRmatrix}
R=\frac{1}{2d^2}
\begin{pmatrix}
1\!+\!q\bar{q}\!-\!h\bar{h}&\hspace{-.3cm}0&\hspace{-.3cm}0&\hspace{-.3cm}0&\hspace{-.3cm}0&\hspace{-.3cm}\!-\!2\bar{h}&\hspace{-.3cm}0&\hspace{-.3cm}2\bar{h}q&\hspace{-.3cm}0\cr 
0&\hspace{-.3cm}2\bar{q}&\hspace{-.3cm}0&\hspace{-.3cm}1\!-\!q\bar{q}\!+\!h\bar{h}&\hspace{-.3cm}0&\hspace{-.3cm}0&\hspace{-.3cm}0&\hspace{-.3cm}0&\hspace{-.3cm}2 h\bar{q}\cr 
0&\hspace{-.3cm}0&\hspace{-.3cm}2q&\hspace{-.3cm}0&\hspace{-.3cm}\!-\!2h&\hspace{-.3cm}0&\hspace{-.3cm}q\bar{q}\!+\!h\bar{h}\!-\!1&\hspace{-.3cm}0&\hspace{-.3cm}0\cr 
0&\hspace{-.3cm} q\bar{q}\!+\!h\bar{h}\!-\!1&\hspace{-.3cm}0&\hspace{-.3cm}2q&\hspace{-.3cm}0&\hspace{-.3cm}0&\hspace{-.3cm}0&\hspace{-.3cm}0&\hspace{-.3cm}\!-\!2h\cr 
0&\hspace{-.3cm}0&\hspace{-.3cm}2\bar{h}q&\hspace{-.3cm}0&\hspace{-.3cm}1\!+\! q\bar{q}\!-\!h\bar{h}&\hspace{-.3cm}0&\hspace{-.3cm}\!-\!2\bar{h}&\hspace{-.3cm}0&\hspace{-.3cm}0\cr 
2h\bar{q}&\hspace{-.3cm}0&\hspace{-.3cm}0&\hspace{-.3cm}0&\hspace{-.3cm}0&\hspace{-.3cm}2 \bar{q}&\hspace{-.3cm}0&\hspace{-.3cm}1\!-\!q\bar{q}\!+\!h\bar{h}&\hspace{-.3cm}0\cr 
0&\hspace{-.3cm}0&\hspace{-.3cm}1\!-\!q \bar{q}\!+\!h\bar{h}&\hspace{-.3cm}0&\hspace{-.3cm}2h\bar{q}&\hspace{-.3cm}0&\hspace{-.3cm}2\bar{q}&\hspace{-.3cm}0&\hspace{-.3cm}0\cr 
\!-\!2 h&\hspace{-.3cm}0&\hspace{-.3cm}0&\hspace{-.3cm}0&\hspace{-.3cm}0&\hspace{-.3cm}q\bar{q}\!+\!h\bar{h}\!-\!1&\hspace{-.3cm}0&\hspace{-.3cm}2q&\hspace{-.3cm}0\cr
 0&\hspace{-.3cm}\!-\!2\bar{h}&\hspace{-.3cm}0&\hspace{-.3cm}2\bar{h}q&\hspace{-.3cm}0&\hspace{-.3cm}0&\hspace{-.3cm}0&\hspace{-.3cm}0&\hspace{-.3cm}1\!+\!q\bar{q}\!-\!h\bar{h}\cr 
\end{pmatrix}
\ee
\normalsize
The first observation about the above $R$--matrix is that is cyclic,
$R^{ab}_{cd}=R^{(a+1)(b+1)}_{(c+1)(d+1)}\!=\!R^{(a\!-\!1)(b\!-\!1)}_{(c\!-\!1)(d\!-\!1)}$. This feature, 
which can be traced back to the cyclic quantum plane relations (\ref{cyclicplane}) 
(which in turn was forced upon us by the need to preserve the cyclic $\Zset_3$ symmetry)
distinguishes this $R$--matrix from those corresponding to standard quantum deformations of $\SU(3)$,
see e.g. \cite{Majid, EwenOgievetsky94}. Those are related to the symmetries of \emph{ordered}
Manin planes and are thus not cyclic. 

 It is also straightforward to check that this $R$--matrix leads to the expression (cf. (\ref{ch3:forminvariance}))
\be
f(x^a,x^{a+1})=\R ak{\!\!a+1}{\!\!l} x^kx^l-x^{a+1}x^a=\left( x^ax^{a+1}-q x^{a+1} x^a+ h x^{a-1}_{a-1} x^{a-1}_{a-1}\right)\cdot\qb/d^2
\ee
with consistent relations from $f(x^{a+1}\!,x^a)$ and $f(x^a\!,x^a)$ and similarly for the co--plane coordinates. Note that
$\hat{R}$ has $1$ as one of its eigenvalues, so we chose $\lambda\!=\!1$ in defining the quantum plane (cf. (\ref{vector})). 
 Setting $f(x^a,x^{a+1})\!=\!0$ we thus reproduce the cyclic quantum plane structure in (\ref{cyclicplane}). 
However, according to the general discussion in section 3, and as will be discussed more thoroughly in the 
following, the quantum algebra produced by $R$ leaves not just $f(x^a,x^{a+1})\!=\!0$ invariant, but the form of the full 
``off shell'' expression $f(x^a,x^{a+1})$. It will thus lead to a symmetry of the Lagrangian 
itself and not only of the moduli space.

 The final important property of (\ref{qhRmatrix}) is that, for generic values of the deformation
parameters, it does not satisfy the Yang--Baxter equation. It is thus a slight abuse of language
to call it an $R$--matrix (to emphasise this, we will sometimes refer to it as a \emph{generalised}
$R$--matrix). The fact that the YBE is not satisfied means that it is not automatic that the
RTT relations have nontrivial solutions, and furthermore the associativity of the algebra will
lead to new relations at cubic order.

\subsubsection{The quadratic algebra relations}

The quantum algebra relations following from the $R$--matrix in (\ref{qhRmatrix}) are derived in 
more detail in Appendices B and C. Here we will just tabulate the resulting
independent quadratic relations in Table \ref{algebrarelations}. 

\begin{table}[h]
\begin{center}
\begin{tabular}{|cc|} \hline 
&  \\ 
(a)&$ \tq ac t^{a+1}_{\;c}-q t^{a+1}_{\; c}\tq ac +ht^{a-1}_{\;c}t^{a-1}_{\;c}=
h\left(t^a_{\; c+1} t^{a+1}_{\;c-1}
-\bar{q}t^a_{\; c-1} t^{a+1}_{\;c+1}+\bar{h} t^a_{\; c} t^{a+1}_{\;c}\right)$\\ 
&\\ 
(b)& $q[\tq{a+1}{c+1},\tq a c]=-q^2\tq{a+1} c\tq a {c+1}+hq \tq {a-1}c\tq{a-1}{c+1}
+h\tq{a-1}{c+1}\tq{a-1}c+\tq a{c+1}\tq{a+1}c$\\
&\\ 
(c)&$-qt^{a+1}_{\; c}t^a_{\; c+1}+\bar{q}t^{a}_{\; c+1}t^{a+1}_{\; c} =
\bar{h} t^a_{\; c-1} t^{a+1}_{\;c-1}
-h t^{a-1}_{\;c}t^{a-1}_{\;c+1} $\\ 
& \\
(d)&$h(t^{a}_{\; c+1}t^a_{\; c-1}-\bar{q} t^{a}_{\; c-1}t^{a}_{\; c+1}) =
\bar{h}(t^{a+1}_{\; c} t^{a-1}_{\;c}
-q t^{a-1}_{\;c}t^{a+1}_{\;c}) $\\ 
&\\ \hline
\end{tabular}
\caption{The quadratic quantum algebra relations for the algebra $\Acal(R)$ corresponding to the 
general $(q,h)$--deformation. The indices are identified modulo 3, e.g. $a+1=a-2$. }  \label{algebrarelations} 
\end{center}
\end{table}
We will refer to the quantum algebra defined by (\ref{Hsymmetry}) as $\Acal(R)$. 
There are several crucial things which have to work for $\Acal(R)$ to be a
consistent algebra. First of all, it is highly non-trivial to have a solution in general
to the equation (\ref{Hsymmetry}). We already know that for general
values of $h$ and $q$ there is no $R$--matrix related to the quantum 
plane which satisfies the YBE (had there been, we would be 
guaranteed a non-trivial solution to (\ref{Hsymmetry})). Luckily, 
in the table we have exactly 36 independent relations, which is what is needed.
As shown in appendix B, all remaining relations are linearly dependent on these. In appendix
D, we discuss the possibility of representing the elements $\tq ij$ satisfying these relations
in terms of matrices. 

 The Yang-Baxter equation also ensures that the associativity of the algebra 
does not lead to additional cubic relations. When $R$ does not belong to the equivalence 
class of $R$--matrices satisfying the  YBE, the ideal generated by the quadratic relations 
will contain higher order (at least cubic) terms. This leads to a potential danger, since
the ideal could in principle become too large and trivialise the algebra at cubic order.  
The cubic relations are analysed in Appendix C, with the result that the new relations
do not spoil the desired properties of the algebra, like the existence of an antipode and
quantum determinant, as we will now discuss.

\subsubsection{The antipode and quantum determinant}

 As discussed, for $\Acal(R)$ to be a symmetry of the Lagrangian the quantum determinant (\ref{ch3:QDdef}) 
has to be central. Thanks to the cubic relations derived in appendix C, we will manage to prove that this
is the case. In the process we will show the existence of an antipode, thus showing that our bialgebra
is in fact a Hopf algebra. Let us explicitly write out the quantum determinant for the case under consideration: 
\small
\be
\label{FullQuantumdeterminant}
\begin{split}
\mathbb{D}=&\frac1{6d^2}\left( \right. 
\tq 11 \tq 22 \tq 33\!-\!q\tq 21 \tq 12 \tq 33+h\tq 31 \tq 32 \tq 33
+\tq 31 \tq 12 \tq 23\!-\!q\tq 11 \tq 32 \tq 23+h\tq 21 \tq 22 \tq 23
+\tq 21 \tq 32 \tq 13\!-\!q\tq 31 \tq 22 \tq 13+h\tq 11 \tq 12 \tq 13 \\
&\!-\!\bar{q}( \tq 11 \tq 23 \tq 32\!-\!q\tq 21 \tq 13 \tq 32+h\tq 31 \tq 33\tq 32
+\tq 31 \tq 13 \tq 22\!-\!q\tq 11 \tq 33\tq 22+h\tq 21 \tq 23 \tq 22
+\tq 21 \tq 33 \tq 12\!-\!q\tq 31 \tq 23 \tq 12+h\tq 11 \tq 13 \tq 12)\\
&+\bar{h}(\tq 11 \tq 21 \tq 31\!-\!q\tq 21 \tq 11 \tq 31+h\tq 31 \tq 31 \tq 31
+\tq 31 \tq 11 \tq 21\!-\!q\tq 11 \tq 31 \tq 21+h\tq 21 \tq 21 \tq 21
+\tq 21 \tq 31 \tq 11\!-\!q\tq 31 \tq 21 \tq 11+h\tq 11 \tq 11 \tq 11)\\
&+\mbox{cyclic permutations}\left. \right) \;.
\end{split}
\ee
\normalsize
It is possible to see that all the rows above are proportional to each other 
just directly from the quadratic relations,
which we will demonstrate with the first two rows. Below we write what 
the first row minus the second row times $(q\bar{q})^{-1}$ is
\be
\begin{split}
&\{\mbox{row one}\}-(q\bar{q})^{-1}\{\mbox{row two}\}=\\
=&\tq 11 (\tq 22 \tq 33-\tq 33\tq 22 -q\tq 32\tq 23+q^{-1}\tq 23\tq 32
 +h\tq 12 \tq 13-hq^{-1}\tq 13\tq 12)\\
&+\tq 21 (-q\tq 12 \tq 33-\tq 13\tq 32+h\tq 22\tq 23-hq^{-1}\tq 23\tq 22
+\tq 32\tq 13+q^{-1}\tq 33\tq 12)\\
&+\tq 31 (h\tq 32\tq 33+hq^{-1}\tq 33\tq 32+\tq 12\tq 23+q^{-1}\tq 13\tq 22-
q\tq 22\tq 13-\tq 23\tq 12 )=0\;.
\end{split}
\ee
Every parenthesis above is separately zero as a consequence of the $(b)$
quadratic relation in Table \ref{algebrarelations}. 
Note that this is consistent with the cubic relations from equation (\ref{ch4:invariance1}).
There is also a symmetry for these relations if we interchange upper and lower indices and
at the same time let $(\bar{q},\bar{h})\rightarrow (q,h)$. 

The next step is to show that all cyclic permutations of the first row are equal. This is 
again straightforward. Let us write one of them:
\be
\underline{\tq 22 \tq 33 \tq 11}-q\tq 32 \tq 23 \tq 11+h\tq 12 \tq 13 \tq 11
+\underline{\tq 12 \tq 23 \tq 31}-q\tq 22 \tq 13 \tq 31+h\tq 32 \tq 33 \tq 31
+\underline{\tq 32 \tq 13 \tq 21}-q\tq 13 \tq 33 \tq 21+h\tq 22 \tq 23 \tq 21\;.
\ee
Now move the last ${\bf t}$ factor in each of the underlined terms to the front, commuting
it through the other two ${\bf t}$'s. Again for this only relation $(b)$ is needed. All unwanted terms  
cancel and we are left precisely with the first row in (\ref{FullQuantumdeterminant}).
Proceeding in this way we conclude that
\be \label{FinalQuantumDeterminant}
\mathbb{D}=\tq 11 \tq 22 \tq 33-q\tq 21 \tq 12 \tq 33+h\tq 31 \tq 32 \tq 33
+\tq 31 \tq 12 \tq 23-q\tq 11 \tq 32 \tq 23+h\tq 21 \tq 22 \tq 23
+\tq 21 \tq 32 \tq 13-q\tq 31 \tq 22 \tq 13+h\tq 11 \tq 12 \tq 13.
\ee
It is crucial that the cubic constraints discussed in appendix \ref{cubicappendix} can never take
this particular form. So $\mathbb{D}$ is guaranteed to be a nontrivial cubic element. We now have
to show that it is central. 

 Centrality of the quantum determinant is related to the existence of an antipode ${\bf s}$, which 
can be thought of as an inverse matrix to $\tq ij$ satisfying
\be
\label{antipode}
t^i_{\,k}s^k_{\,j}=\delta^i_{\,j}
\qquad \mbox{and} \qquad
s^i_{\,k}t^k_{\,j}=\delta^i_{\,j}\,.
\ee
This is because, if it is possible to find a matrix ${\bf s}$ which  satisfies
\be
\label{almost antipode}
t^i_{\,k}s^k_{\,j}=\delta^i_{\,j}\mathbb{D}
\qquad \mbox{and} \qquad
s^i_{\,k}t^k_{\,j}=\delta^i_{\,j}\mathbb{D}\,,
\ee
it follows that $\mathbb{D}$ is central\footnote{Multiplying the first of (\ref{almost antipode})
by $\tq jl$ on the right, we find $\tq ik \sq kj \tq jl=\mathbb{D}\delta^i_{\,j}\tq jl
\implies \tq ik \delta^k_{\; l}\mathbb{D}=\mathbb{D}\, \tq il
\implies \tq il \mathbb{D}=\mathbb{D}\, \tq il$. }
 and therefore can be chosen to equal one, and  ${\bf s}$ would satisfy (\ref{antipode}).  
We will now check that the  following ${\bf s}$ satisfies (\ref{almost antipode}):
\be
\label{ch4:Antipodegeneral}
s^{1+i}_{\;1+k}=t^{2+k}_{\;2+i}t^{3+k}_{\;3+i}
-\bar{q} t^{2+k}_{\;3+i}t^{3+k}_{\;2+i}+
\bar{h} t^{2+k}_{\;1+i}t^{3+k}_{\;1+i}=
t^{2+k}_{\;2+i}t^{3+k}_{\;3+i}-q t^{3+k}_{\;2+i}t^{2+k}_{\;3+i}+h t^{1+k}_{\;2+i}t^{1+k}_{\;3+i}\,.
\ee
The two expressions for ${\bf s}$ are easily seen to be equal by use of relation $(c)$ in Table 1. 

First we check the diagonal elements. Writing out e.g. $\sq 3k \tq k3$ we have
\be
\sq 3k \tq k3=
\tq 11 \tq 22 \tq 33-q\tq 21 \tq 12 \tq 33+h\tq 31 \tq 32 \tq 33
+\tq 31 \tq 12 \tq 23-q\tq 11 \tq 32 \tq 23+h\tq 21 \tq 22 \tq 23
+\tq 21 \tq 32 \tq 13-q\tq 31 \tq 22 \tq 13+h\tq 11 \tq 12 \tq 13
\ee
which is nothing but $\mathbb{D}$ in (\ref{FinalQuantumDeterminant}). 
The same can be checked for the other two diagonal elements, which can be 
obtained from cyclicity. We have thus shown that the diagonal terms in ${\bf st}$ are
all proportional to the quantum determinant. 

 To see the vanishing of the off--diagonal terms, we will need
to employ the nontrivial cubic relations in appendix \ref{cubicappendix}.
For instance let us write explicitly the off--diagonal term
\be
\label{ch4:offdiagonal1}
\begin{split}
\sq 3k \tq k2=
&\tq 11 \tq 22 \tq 32\!-\!q\tq 21 \tq 12 \tq 32\!+\!h\tq 31 \tq 32 \tq 32
\!+\!\tq 31 \tq 12 \tq 22\!-\!q\tq 11 \tq 32 \tq 22\!+\!h\tq 21 \tq 22 \tq 22
\!+\!\tq 21 \tq 32 \tq 12\!-\!q\tq 31 \tq 22 \tq 12\!+\!h\tq 11 \tq 12 \tq 12 \\
=& \tq 11(\tq 22\tq 32-q \tq 32 \tq 22+h \tq 12 \tq 12)
+\tq 21( \tq 32 \tq 12-q \tq 12 \tq 32+h \tq 22 \tq 22)
+\tq 31 (\tq 12 \tq 22-q \tq 22 \tq 12+h \tq 32 \tq 32)\;.
\end{split}
\ee
That this vanishes follows from the cubic constraint (\ref{L122conj}).

In order to complete the proof, we have to check that we get the same if we multiply
${\bf t}$ and ${\bf s}$ in the reverse ordering.
Now we use the second expression for the components $\sq kl$, to compute e.g.
\be\label{ch4:diagonal2}
\tq 3k \sq k3=
\tq 11 \tq 22 \tq 33-\bar{q}\tq 12 \tq 21 \tq 33+\bar{h}\tq 13 \tq 23 \tq 33
+\tq 13 \tq 21 \tq 32-\bar{q}\tq 11 \tq 23 \tq 32+\bar{h}\tq 12 \tq 22 \tq 32
+\tq 12 \tq 23 \tq 31-\bar{q}\tq 13 \tq 22 \tq 31+\bar{h}\tq 11 \tq 21 \tq 31\, ,
\ee
which is also equal to $\mathbb{D}$ using relation $(c)$. As for the off--diagonal terms,
we can check e.g.
\be
\label{ch4:offdiagonal2}
\tq 1k \sq k3=
(\tq 12 \tq 13 -\bar{q}\tq 13 \tq 12 +\bar{h}\tq 11 \tq 11) \tq 21
+(\tq 13 \tq 11 -\bar{q}\tq 11 \tq 13 +\bar{h}\tq 12 \tq 12 )\tq 22
+(\tq 11 \tq 12 -\bar{q}\tq 12 \tq 11 +\bar{h}\tq 13 \tq 13) \tq 23
\, .
\ee
This is zero due to the cubic constraint (\ref{L112}) following from the RTT relations (see Appendix C).

 It is crucial that the cubic relations relating the diagonal terms come from the components of the 
YBE which are fulfilled (these are the choices of indices where no new cubic constraints arise),  
while the off--diagonal relations arise when the YBE is not 
fulfilled. If the YBE had not been satisfied for the diagonal relations they would have become zero
and we would not have been able to define the quantum determinant. 

We conclude that our matrix quantum algebra $\Acal(R)$ is equipped with an antipode and is thus
a Hopf algebra. Furthermore, the quantum determinant is central, which, as discussed above, implies
that we can set $\mathbb{D}=1$. This makes the superpotential $\Acal(R)$--invariant.

\subsubsection{Invariance of the full Lagrangian} \label{fullinv}

Up to this stage we have only been discussing the invariance of the F-terms in the Lagrangian
under $\Acal(R)$. Having shown that there exists an antipode, we can now check invariance of the 
kinetic term $\Tr \Phibar e^V\Phi e^{-V}$. Since the $e^V$'s, being $\SU(3)$ singlets, are not relevant for 
this, we can simply check invariance of $\Phibar \Phi$:
\be
\Phibar_i \Phi^i\rightarrow \Phibar_j {t^j_{\;\; i}}^* \tq ik \Phi^k \;.
\ee
Thus invariance requires ${t^j_{\;\; i}}^* \tq ik=\delta^j_{\;\;k}$, which
of course is satisfied because  ${\bf t}^*$ is  the antipode
(\ref{ch4:def.antipode}). Note that the reality condition on the $R$--matrix, which 
reduced the quantum deformation of $\SL(3)$ down to a quantum deformation of $\SU(3)$, is
crucial for this to hold. 

 As a consequence of $\Acal(R)$--invariance of the scalar kinetic term we conclude that the D--terms
(and therefore the full Lagrangian) are also invariant. 

Quantum groups have long been known to play a fundamental role in two--dimensional physics, and 
2d conformal field theory in particular \cite{Gomezetal}. So far, their role in four--dimensional field
theory has been much more limited, although they have been considered both as candidates for gauge
groups \cite{Sudbery96} (see also \cite{Mesref04} for a recent review and references) and flavour groups
(see e.g. \cite{Gavrilik01} for a summary of results in this direction). We have just constructed a new example: 
Being a quantum deformation of the $\SU(3)$ R--symmetry group, $\Acal(R)$ plays the role of a flavour group in 
the Leigh--Strassler theories (though the flavours here are adjoint).

\subsection{Quantum symmetry and finiteness?} \label{qsandfiniteness}

In the preceding sections we showed that the classical $(q,h)$--deformed Leigh--Strassler 
Lagrangian enjoys a Hopf algebra symmetry $\Acal(R)$, which is defined through the $R$--matrix
(\ref{qhRmatrix}) related in a simple way to the holomorphic one--loop spin chain Hamiltonian of the
theory. Since the classical Lagrangian knows nothing about spin chains, this should be 
understood in the opposite direction: That the one--loop Hamiltonian has $\Acal(R)$ as a symmetry 
is a consequence of $\Acal(R)$ not being broken at one--loop level in the planar limit. 
It should be emphasised that we have not been assuming the planar limit in the discussion
above\footnote{Among other things, this indicates that the full dilatation operator of the theory (i.e. 
including non--planar corrections) should exhibit the quantum symmetry. Confirming this would provide
a non--trivial check of our construction.}.

So is $\Acal(R)$ the hidden symmetry that, as argued in the introduction, might be related to the
finiteness properties of the Leigh--Strassler theories? The story is certain to be more subtle, since
the finiteness condition (\ref{finiteness}) depends non--trivially on the number of colours $N$, while 
(as just discussed) $\Acal(R)$ was defined without any reference to $N$. Assuming that there is a 
correlation, it could be that the requirements of finiteness and quantum symmetry invariance only overlap 
in the planar limit. The fact that, as we saw, the most natural normalisation matches what is required by
planar finiteness points in this direction. This normalisation is forced upon us if we wish
to impose (\ref{ch3:normalisationEF}) as part of our definitions. That condition was crucial 
for \cite{EwenOgievetsky94} who were working in the quasi--triangular case, but since we are relaxing 
several of their assumptions we do not yet have an argument for why (\ref{ch3:normalisationEF}), with
the precise factor of $1/2$, is singled out from the matrix quantum algebra point of view. 
This link definitely deserves to be explored further. 

Conversely, it might be that the definition of $\Acal(R)$ could be suitably extended 
to involve $N$ so as to make contact with the full finiteness condition (\ref{finiteness}). 
It is also possible that finiteness and $\Acal(R)$ are completely unrelated,
which would be demonstrated most clearly by finding an example of a non--finite theory with a similar
quantum symmetry structure. Since a simple way of keeping $\Acal(R)$ without preserving finiteness would
be to (non--supersymmetrically) change the relative coefficients between the $F$-- and $D$--terms, 
it will be important to go beyond the holomorphic sector and include the $D$--terms in the discussion 
of $\Acal(R)$.

Leaving the resolution of these issues to future work, we will continue to explore the properties
of our novel quantum symmetry algebra by considering how it is acted upon by the known discrete symmetries 
of the Lagrangian.

\subsection{Discrete symmetry in the general deformation} \label{Discrete}

As discussed above, the general Leigh--Strassler deformations preserve certain  discrete
symmetries: A cyclic $\Zset_3$ symmetry and another $\Zset_3$ which acts by multiplying the
scalars by a third root of unity. Acting on the three scalar superfields, these symmetries can be 
represented as shift and clock matrices \cite{Aharonyetal02}:
\be \label{shiftclock}
U=\threebythree{0}{1}{0}{0}{0}{1}{1}{0}{0}\,,\quad V=\threebythree{1}{0}{0}{0}{\omega}{0}{0}{0}{\omega^2}\;.
\ee
This representation makes it clear that these are (very special) elements of the original $\SU(3)$ 
subgroup of the $\Ncal=4$
R--symmetry which was acting on the scalars. In addition to these, there is another, central, $\Zset_3$ symmetry within
$\Urm(1)_R$ which, in this basis, simply acts as $W=\omega 1\!\!1$. All these $\Zset_3$'s do not commute, rather they 
combine to produce a trihedral group with 27 elements (given by all combinations of the generators $U$, $V$ and $W$ 
up to the relations $U^3=V^3=W^3=1$ and $UV=WVU$) known as $\Delta_{27}$ 
\cite{Aharonyetal02,Wijnholt05}. Some aspects of this
discrete group have been investigated in \cite{MadhuGovindarajan07a,MadhuGovindarajan07b}, where it was shown that
it is unbroken at the first few orders in perturbation theory. We should emphasise that the fact that these symmetries
are preserved at the quantum level is a crucial consistency check, since in particular the cyclic symmetry is used
as input in the Leigh--Strassler argument which equates the anomalous dimensions of the three scalars. 

Given our new--found understanding of the general Leigh--Strassler deformation as a quantum deformation of $\SU(3)$, 
rather than a simple breaking to $\Zset_3\times\Zset_3$, it is important to clarify the role of the discrete symmetries
in our setting. As we will now show, they simply act as automorphisms of the quantum symmetry algebra. Given
the close relationship of the general Leigh--Strassler deformation to cubic forms \cite{EwenOgievetsky94,Wijnholt05} 
this is of course not surprising. In the following
we will closely follow the discussion in \cite{EwenOgievetsky94} for the deformations of $\GL(3)$ they consider.

Recall that, for the quantum plane, an automorphism is a mapping $x^i\ra Z^i_{\;j} x^j$ 
which leaves $E_{ijk}$ invariant (and similarly for the co--plane). For the $E_{ijk}$ corresponding
to the general deformation we can easily check invariance under the above transformations.
We are now interested in how the automorphism group acts on the algebra generators. This will be by
conjugation, as $\tq ij \ra Z^i_{\;k}\tq k l {Z^{-1}}^l_{\; j}$. We thus find:
\be
U: \quad \threebythree{\tq 11}{\tq 12}{\tq 13}{\tq 21}{\tq 22}{\tq 23}{\tq 31}{\tq 32}{\tq33}
\longrightarrow  \threebythree{\tq 22}{\tq 23}{\tq 21}{\tq 32}{\tq 33}{\tq 31}{\tq 12}{\tq 13}{\tq 11}\;, \quad\text{i.e.}
\quad \tq ab\ra \tq {a+1} {b+1}
\ee
while 
\be
V: \quad  \threebythree{\tq 11}{\tq 12}{\tq 13}{\tq 21}{\tq 22}{\tq 23}{\tq 31}{\tq 32}{\tq33}
\longrightarrow \threebythree{\tq 11}{\omega^2\tq 12}{\omega\tq 13}{\omega\tq 21}{\tq 22}{\omega^2\tq 23}
{\omega^2\tq 31}{\omega\tq 32}{\tq33}\;, \quad 
\text{i.e.} \quad \left\{\begin{array}{l}\tq aa \ra \tq aa\\
\tq {a+1}a\ra \omega\tq{a+1}a\\ \tq a{a+1}\ra \omega^2\tq a{a+1}\end{array}\right. \;. 
\ee
It can be easily checked that the algebra commutation relations tabulated in Table \ref{algebrarelations} are 
invariant under these transformations, as well as their combinations. So we have explicitly
exhibited how the discrete symmetries act on the quantum symmetry algebra $\mathcal{A}(R)$.

It might be of interest to note that the elements $U$ and $V$ can be obtained by suitable
truncations of the full $\Acal(R)$ algebra. 
The action of $U$ on the scalar field is:
\be \label{Z31t}
\Phi^1\rightarrow \tq 12\Phi^2\quad,\quad \Phi^2\rightarrow \tq 23 \Phi^3\quad,\quad \Phi^3\rightarrow \tq31 \Phi^1\;.
\ee
So to exhibit this symmetry, we truncate the algebra by setting $\tq ij=0$ except 
for $\tq 12,\tq 23,\tq 31$. Looking at the relations in Table \ref{algebrarelations}, 
we see that the only nontrivial ones left are 
\be \label{t1223com}
[\tq 12,\tq 23]=0 \qquad \big(\text{from} \quad q[\tq 23,\tq12]=\tq 13\tq22-q^2\tq22\tq13+hq \tq32\tq33+h\tq33\tq32\big) 
\ee
and
\be \label{t122331}
\tq 12\tq 23=(\tq 31)^2 \qquad \big(\text{from} \; 
\tq11\tq21-q\tq21\tq11+h\tq31\tq31=h\left(\tq12\tq23-\qb\tq13\tq22+\hb\tq11\tq21\right)\big)
\ee
as well as their cyclic relations. From (\ref{t1223com}) we conclude that in this subsector the $\tq a{a+1}$ commute among
themselves, so that we can treat them as actual numbers. From the constraint (\ref{t122331}) we then conclude that they
have to be cubic roots of unity, and in the simplest case (corresponding to $U$) they can be set to 1. 

As for the action of $V$, it is 
\be \label{Z32t}
\Phi^1\rightarrow \tq11\Phi^1\quad,\quad \Phi^2\rightarrow\tq22\Phi^2\quad,\quad \Phi^3\rightarrow \tq33\Phi^3
\ee
where again we need to show that when all other $\tq ij$ are set to zero, the three $\tq aa$ commute.
This is also the case:
\be
[\tq11,\tq22]=0 \quad \big(\text{from} \; q[\tq11,\tq22]=-\tq12\tq21+q^2\tq21\tq12-hq\tq31\tq32-h\tq32\tq31\big)
\ee
and 
\be
\tq11\tq22=(\tq33)^2 \quad \big(\text{from} \; 
\tq13\tq23-q\tq23\tq13+h\tq33\tq33=h\left(\tq11\tq22-\qb\tq12\tq21+\hb\tq13\tq23\right)\big)
\ee
and cyclic permutations. We can thus choose $\tq 11=1$, $\tq 22=\omega$ and $\tq33=\omega^2$, obtaining the element $V$. 

There is yet a third class of elements obtained by setting all generators to zero apart from $\tq {a+1}a$, and 
proceeding in this way we can see how the remaining elements of $\Delta_{27}$ can be embedded in the quantum
symmetry algebra.\footnote{Note that (as per the discussion below (\ref{relations:1})) successive actions of ${\bf t}$ belong
to different copies of $M_n$ and thus commute with each other.}

\subsection{Beyond the $\SU(3)$ sector}

Up to now we have been mostly interested in the quantum symmetry underlying the holomorphic sector
of the theory, spanned by the three scalar superfields $\Phi^i$ that enter the superpotential. 
As discussed, these can be usefully mapped to the coordinates $x^i$ of a quantum plane defined by the
$R$--matrix (\ref{qhRmatrix}) through the relations (note that, as mentioned, $1$ is an eigenvalue of $\hat{R}$, 
which allows us to choose $\lambda=1$ in (\ref{vector}))
\be 
\R ikjl x^k x^l=x^jx^i
\ee
with the RTT relations (\ref{RTT}) guaranteeing invariance of the
plane under the quantum symmetry transformations $x^i\ra \tq ij x^j$. 

Similarly the antiholomorphic scalars are mapped to the coordinates $\bar{x}^{\bar{i}}=u_i$ of a quantum co--plane,
 defined through the same $R$--matrix via 
\be \label{Ruu}
u_k u_l \R kilj=u_j u_i
\ee
Since, as discussed, the co--plane coordinates transform as $u_i\ra u_j \sq ji$ under the quantum symmetry (where
$\mathrm{s}$ is the antipode), the relation that guarantees invariance of the co--plane reads 
$\sq ri\sq sj\R ikjl=\R risj \sq jl \sq ik$. This can be easily seen to follow from the original RTT relations. 
 
 However, a moment's thought shows that this cannot be the end of the story. The full Hamiltonian certainly 
mixes the plane and co--plane coordinates, which means that the ($36\times 36$) $R$--matrix which should define
the quantum symmetry of the full scalar field sector will also imply nontrivial commutation relations between
the $x^i$ and $u_i$ planes. If we converted to real notation (schematically $y^I=x^i\pm i u_i$, $I=1\ldots 6$)
we would get a six dimensional quantum plane acted on by a suitable deformation of $\mathrm{SO}(6)$ \cite{BerensteinCherkis04}.

On the other hand, since the Leigh--Strassler theories arise just through a
superpotential deformation, there should not be any additional information in the $36\times 36$ $R$--matrix than 
that which is already contained in the holomorphic $9\times9$ $R$--matrix (\ref{qhRmatrix}). So the mixed 
commutation relations should be derivable through suitable conditions involving $R$. 

 Following e.g. \cite{Majid} (see also \cite{WessZumino91}), we propose the following 
as suitable definitions for the mixed planes:

\be \label{MixedPlane}
u_l \R jkli x^k=x^j u_i\; \quad \text{and}\quad x^k \Rt iklj u_l=u_j x^i\;.
\ee
Here $\tilde{R}$ is the so--called \emph{second inverse} of R, defined through\footnote{The normal inverse
of $R$ of course simply satisfies $(R^{-1})^{i\; j}_{\;m\; n}R^{m\;n}_{\;\;k\;\;l}=\delta^i_{\;k}\delta^j_{\;l}=
R^{i\;\; j}_{\;m\; n}(R^{-1})^{m\;n}_{\;\;k\;\;l}.$}
\be
\Rt imnj \R mlkn=\delta^i_{\;l}\delta^k_{\;j}=\R imnj \Rt mlkn\;.
\ee

These relations are invariant under the quantum symmetry transformations of $x^i$ and $u_i$. To see this, 
one needs to use the relations 
\be 
\sq sj \R akjl \tq kb =\tq ak \R kbsj \sq jl\;,\quad \text{and}\quad
\sq sd \Rt befs \tq cb=\tq ae \Rt cald \sq fl
\ee
which also follow straightforwardly from the original RTT relations. 

The above mixed plane relations will be useful in section \ref{Noncommutative} where we will make contact with
previous work on the noncommutative description of the Leigh--Strassler theories.

\section{The integrable cases} \label{IntegrableCases}

 Having discussed the general framework necessary to understand the quantum symmetry of the general 
Leigh--Strassler deformation, in this section we will focus on the subset of cases which are 
integrable. Not surprisingly, these are the special cases where the $R$--matrix
(\ref{qhRmatrix}) satisfies the YBE, and $\Acal(R)$ becomes dual to a quasi--triangular Hopf algebra. 
While the material in this section is not new, we believe that it is appealing and instructive to 
reconsider these cases from the Hopf algebra perspective we have been developing.

\subsection{The real $\beta$ deformation}

 As the simplest example of the general discussion above, we turn to the case where $h=0$, while $q$
is taken to be just a phase, $q=e^{i\beta}$, with $\beta$ real. As mentioned previously, this is a well known
case where integrability of the one--loop dilatation operator of $\Ncal=4$ is preserved. 
So it is to be expected that this case will reveal even more structure
than the general $(q,h)$--deformation. Due to the amount of attention this particular case has received
in the literature, we will aim to keep the discussion in this section self--consistent, at the expense
of some repetition from the previous section. 

In this case the superpotential is simply given by 
\be
\Wcal=\Tr[\Phi^1\Phi^2\Phi^3-q \Phi^1\Phi^3\Phi^2]
\ee
which leads us to the following choice for $E_{ijk}$, $F^{ijk}$:
\be
E_{123}=1\;, \; E_{132}=-q\; ,\; F^{123}=1\; ,\; F^{132}=-\frac1{q} \quad \big(\text{+ cyclic permutations}\big)\;.
\ee
According to the above general discussion, the quantum algebra transformations $\Phi^i \rightarrow \tq ij \Phi^j$ that
leave the superpotential invariant will be generated by an $R$--matrix constructed through $E_{ijk}$ and $F^{ijk}$.

\mparagraph{The $R$--matrix for $\beta$ real}

As before, this $R$--matrix will be given through $\R ikjl=\Rh jkil$, where $\hat{R}$ is defined by
\be
\Rh ikjl =\delta^i_{\;k}\delta^j_{\;l}- E_{klm}F^{mij}.
\ee
By construction, this $R$--matrix will be cyclic, and is given explicitly by
\be \label{qRmatrix}
R=
\begin{pmatrix}
1&0&0&0&0&0&0&0&0\cr 
0&q^{-1}&0&0&0&0&0&0&0\cr 
0&0&q&0&0&0&0&0&0 \cr 
0&0&0&q&0&0&0&0&0\cr 
0&0&0&0&1&0&0&0&0\cr
0&0&0&0&0&q^{-1}&0&0&0\cr 
0&0&0&0&0&0&q^{-1}&0&0\cr 
0&0&0&0&0&0&0&q&0 \cr 
0&0&0&0&0&0&0&0&1\cr
\end{pmatrix} \;.
\ee
It is worth repeating that this diagonal $R$--matrix is not the one corresponding to the standard 
one--parameter q--deformation of $\SU(3)$ (e.g. \cite{Majid}),
whose $R$--matrix is not cyclic and leads to $i< j$ ordered, rather than cyclic, quantum plane relations. 
However, it is contained as a special case of several $R$--matrices in the literature, such 
as the $X=1$, $q_{13}=1/q_{12}$ case of \cite{Schirrmacher91} and a special
case of the multiparameter $R$--matrix of \cite{Sudbery90}. Diagonal $R$--matrices of this type
have been considered as deformations of $\GL(3)$ in \cite{Zupnik92}. 
 
Furthermore, this $R$--matrix has appeared previously in discussions of integrability for the real $\beta$ 
deformation \cite{BeisertRoiban05}. To be precise, the $R$--matrix presented in that work is more general in 
several respects: It 
applies to more general (nonsupersymmetric) $\gamma$--deformations (where instead of $\beta$ one deforms
by three real phases \cite{Frolov05}), it is valid for the larger $\SU(2|3)$ sector and, importantly, it has
spectral--parameter dependence. Restricting to just real $\beta$ and the $\SU(3)$ sector, the $R$--matrix
of \cite{BeisertRoiban05} is given by\footnote{We have performed a trivial rescaling $\beta\ra -\beta$ 
relative to \cite{BeisertRoiban05}.} 
\be
\R ikjl=\frac{1}{u+i}\left(u e^{-iB_{ij}} \delta^i_{\; k}\delta^j_{\;l}+i P^{i\;\;j}_{\;k\;\;l}\right),
\quad\text{where}\quad
B_{ij}=\threebythree 0 \beta {-\beta} {-\beta} 0 {\beta}{\beta} {-\beta} 0 \;.
\ee
It is easy to check that this $R$--matrix reduces to (\ref{qRmatrix}) in the limit of large
spectral parameter ($u\rightarrow \infty$), which is precisely the regime where one expects to make contact with
the underlying quantum symmetry (for a discussion, see e.g. \cite{Gomezetal}). 
It was noted in \cite{BeisertRoiban05} that the $R$--matrix in this case
was related to the $R$--matrix for the pure $\Ncal=4$ SYM case with
a twist \cite{Reshetikhin90} exactly of the type discussed in section 3, with the nonzero 
matrix elements of $\Fcal$ in (\ref{ch3:twist}) being
\be
\Fcal^{ii}_{\; ii}= 1 \qquad \Fcal^{i i+1}_{\; i i+1} =e^{i\beta/2}
\qquad \Fcal^{i+1 i}_{\; i+1 i}=e^{-i\beta/2} \,.
\ee  
(This is nothing but the $R$--matrix with $\beta\rightarrow -1/2\beta$.)

\mparagraph{The symmetry algebra}

We now turn to the characterisation of the quantum algebra for the real $\beta$ deformation. 
The quantum plane relations one obtains, through $\R ikjl x^k x^l=x^j x^i$, are:
\be
x^1 x^2=q x^2 x^1\;,\quad x^2x^3=q x^3 x^2\;,\quad x^3 x^1=q x^1 x^3 
\ee
which are clearly also cyclic. 

The relations between the various quantum algebra generators following from (\ref{qRmatrix}) (or from the appropriate
limit of those in Table \ref{algebrarelations}) can be summarized as follows:
\be \label{Qalgebraq}
[\tq {a+1}c,\tq a{c-1}]=0\;,\;[\tq a c,\tq {a+1}c]_q=0\;,\;[\tq a c,\tq a {c+1}]_{{q}^{-1}}=0\;,\;[\tq ac,\tq {a+1}{c-1}]_{q^2}=0\;.
\ee
Note that since in this case the $R$--matrix satisfies the Yang--Baxter equation, 
these quadratic relations will not generate additional cubic relations. 
That the $R$--matrix satisfies the YBE implies that the algebra is dual
(in the sense explained in the appendix) to a quasi--triangular Hopf algebra, 
which is the underlying reason that the real $\beta$ deformations are 
integrable.

Using these relations, we can now easily check that the quantum determinant
\be
\mathbb{D}=\tq 11\tq 22\tq33+\tq12\tq23\tq31+\tq13\tq21\tq32-q^{-1}(\tq 11\tq23\tq32+\tq13\tq22\tq31+\tq12\tq21\tq33)
\ee
is central. Setting  $\mathbb{D}=1$ we conclude that the superpotential is indeed invariant under the quantum symmetry.

\mparagraph{The antipode}

Setting $\mathbb D=1$ above reduced the algebra from a $q$--deformation of $\GL(3)$ to a $q$--deformation of 
$\SL(3)$. So far we have not imposed any relations between 
the transformation of the plane and co--plane, or, in other words, a reality condition. From the physical point of
view, it is the invariance of the kinetic term 
which further reduces the algebra down to $\SU(3)$ in the undeformed case. 
On the quantum algebra side, this comes through consideration of
the antipode. By definition, the antipode, when it exists, satisfies 
\be
s^i_{\;j}\tq jk=\tq ij s^j_{\;k}=\delta^i_{\;\;k}\,.
\ee
It is easy to write down the form of the antipode explicitly:
\be \label{Antipode}
S=\begin{pmatrix} 
 \tq 22\tq 33-q^{-1}\tq 23 \tq 32 & \tq 13 \tq 32-q \tq 12\tq 33 & \tq 12\tq23-q^{-1}\tq 13\tq22\\
 \tq 23\tq 31-q^{-1}\tq21\tq33 & \tq 11\tq 33-q\tq13\tq31 & \tq 13\tq 21-q^{-1} \tq 11\tq23 \\
 \tq 21\tq 32-q^{-1}\tq22\tq31 & \tq 12\tq31-q\tq11\tq32 & \tq 11\tq22-q^{-1}\tq 12\tq21 \\
\end{pmatrix} \;.
\ee
Notice this is exactly the antipode (\ref{ch4:Antipodegeneral}) for
$h=0$ and  $\bar{q}=q^{-1}$. We can easily check invariance of the kinetic term as in section
\ref{fullinv}. Thus we have shown that the real--$\beta$ Leigh-Strassler lagrangian is indeed 
invariant under the quantum deformation of $\SU(3)$ provided by $R_\beta$.

\subsection{The other extreme: $q=0, \hb=1/h$} \label{qzerohnonzero}

Let us now turn to other integrable cases which can be embedded in the above framework.
One of the cases considered in \cite{Mansson0703} is that of $q=0$, $\hb=1/h$. 
In this case the epsilon tensor becomes
\be
E_{123}=1\;, \; E_{111}=-h\; ,\; F^{123}=1\; ,\; F^{111}=-\frac1{h} \quad \big(\text{+ cyclic permutations}\big)
\ee
providing the quantum plane relations
\be
x^1 x^2=-h(x^3)^2\quad,\quad x^2 x^3=-h(x^1)^2\quad,\quad x^3 x^1=-h(x^2)^2\;.
\ee
The resulting $R$--matrix is:
\be \label{hRmatrix}
R=\begin{pmatrix}0&0&0&0&0&-{{1}\over{h}}&0&0&0\cr 0&0&0&1&0&0&0&0&0\cr 0&0
 &0&0&-h&0&0&0&0\cr 0&0&0&0&0&0&0&0&-h\cr 0&0&0&0&0&0&-{{1}\over{h}}&0
 &0\cr 0&0&0&0&0&0&0&1&0\cr 0&0&1&0&0&0&0&0&0\cr -h&0&0&0&0&0&0&0&0
 \cr 0&-{{1}\over{h}}&0&0&0&0&0&0&0\cr \end{pmatrix}\;.
\ee
 This $R$--matrix satisfies the quantum Yang--Baxter equation, but it is in several respects rather 
unusual. Taking the undeformed limit $h=1$, we find that it does not reduce to the trivial 
undeformed $R$--matrix (i.e. $1\otimes 1$). As a consequence there does not seem to be a well--defined classical 
r--matrix (see section \ref{rmatrixsection}). Of course there is no reason to expect a 
smooth classical  limit since the point $q=0$ is very special (and very far from the classical point $q=1$). 
We should point out that this $R$--matrix can, in an analogous way as in the real $\beta$-deformed case, be 
obtained from the $R$--matrix with spectral--parameter dependence related to the dilatation operator
found in \cite{BundzikMansson05}.

We should also remark that precisely this choice of parameters ($q=0,\hb=1/h$) has been considered
in the work of \cite{Borketal07} in the context of the finiteness properties of the general Leigh--Strassler
deformation. There it was shown that this is one of only two cases where the 
one--loop (planar) finiteness condition is \emph{exact} to very high loop order (and conjecturally to
all loop orders)\footnote{We wish to thank G. Vartanov for pointing out the relevance of \cite{Borketal07}
to this particular case, as well as M. Kulaxizi for a relevant discussion.}. The other case is the 
real $\beta$ deformation (along with cases which are ``unitary
equivalent'' to real $\beta$ in a sense discussed in \cite{Borketal07}). It would be very interesting to 
further understand the interplay between the quasi--triangular Hopf algebra structure that we have 
exhibited and the higher--loop exactness of the one--loop finiteness condition. 

In \cite{BundzikMansson05} this case was shown to be related to the real $\beta$ case via 
a suitable site--dependent redefinition. Therefore we know that it can be
obtained by a twist starting from the real $\beta$ $R$--matrix of the 
previous section. The matrix 
$\Fcal$ we need is simply the following:
\be
\Fcal=U\otimes U^2
\ee
where $U$ is the shift matrix defined in (\ref{shiftclock}) and where one should rename $q=-1/h$ in 
(\ref{qRmatrix}) to arrive precisely at (\ref{hRmatrix}). Thus the required twist transformation
here is of the generic form  $\Fcal=Z\otimes Z^{-1}$, where  $Z$ is an element of the automorphism group.

\subsection{Other integrable cases and twisting} \label{OtherIntegrable}

There are several other choices for the $R$--matrix (\ref{qhRmatrix}) which solve the Yang--Baxter equation. 
These match the known cases where the general ($q,h$)--deformation gives an integrable one--loop Hamiltonian
in the $\SU(3)$ sector \cite{BundzikMansson05}.
The common characteristic of these solutions is that $q$ and $h$ are related. For instance, the
values  
\be \label{qhrcase}
q=(1+\rho)e^{\frac{2\pi im}{3}}\quad \mbox{and}\quad h=\rho e^{\frac{2\pi i n}{3}}\; 
\quad\qquad (\rho \;\;\text{real}\;,\; m,n\;\; \text{integers})
\ee
lead to integrable Hamiltonians, and from our present point of view to $R$--matrices satisfying the
YBE (as can be easily checked). 
 However, as shown in \cite{BundzikMansson05}  these cases also turned out to 
be related to that of real $\beta$ by similarity transformations,
in some cases combined with site--dependent redefinitions on the spin chain. Similar arguments
based on unitary equivalence were later used in \cite{Borketal07} to demonstrate that they are 
not really new cases.  Also from our Hopf algebra point of view it is straightforward to show that these
cases are related to the real $\beta$ case by Hopf algebra twists. 
The matrix $\Fcal$ in equation (\ref{ch3:twist}) can be implicitly found in 
\cite{BundzikMansson05}. In order to reproduce (\ref{qhrcase}) with the phases set to 
zero, $\Fcal$ takes the form $\Fcal =T  \otimes T$, where $T$ is defined as
\be
T=\frac1{\sqrt{3}}\begin{pmatrix} 1&1&1
\cr 1 & e^{i\frac{2\pi}{3}} & e^{-i\frac{2\pi}{3}}
\cr  1 & e^{-i\frac{2\pi}{3}} & e^{i\frac{2\pi}{3}}\cr
\end{pmatrix} \;,
\ee
and the phase $q$ in (\ref{qRmatrix}) is related to $\rho$ as 
\be
q=\frac{1+2\rho e^{-\frac{\pi i}{3}}+\rho^2e^{-\frac{2\pi i}{3}}}{1+\rho+\rho^2}.
\ee
We see that for zero phases the twist transformation just becomes a similarity transformation. 
This was discussed earlier and corresponds just to a linear basis shift for the generators $\tq ij$.
In order to add the phases in (\ref{qhrcase}) one simply twists again with $\Fcal=Z\otimes Z^{-1}$
as in the previous section, where $Z$ is now taken to be a general element of the automorphism group.

There are certain other parameter choices which were shown to be
integrable in \cite{BundzikMansson05}. They are 
\be
q=-e^{\frac{2\pi i m}{3}}\;\;,\;\; h=e^{\frac{2\pi i n}{3}}
\ee
as well as the very special case $q=0,h=0$. These cases are slightly particular in that
the $R$--matrices arising from directly substituting these values into (\ref{qhRmatrix}) 
do not satisfy the YBE. However, it is easy to verify that they belong to the same equivalence 
class as $R$--matrices that do. To see this, recall from
(\ref{ch3:equivalence}), that the more general definition  $\hat{R}=aI-EF$ leads to 
the same algebra. For the present cases, the values $a=0,2$ produce $R$--matrices 
satisfying the YBE.

 \subsection{Beyond the holomorphic sector?}

 The above are the only known parameter choices where the Leigh--Strassler theories exhibit planar
integrability at one loop, and they could all be seen to easily fit within our formalism, as the
special cases where the quantum algebra $\Acal(R)$ reduces to a quasi--triangular Hopf algebra. 

 However, there is one more case in the literature where one--loop integrability has been observed 
\cite{Mansson0703}. This involves moving out of the holomorphic $\SU(3)$ sector by considering a
sector made up of two holomorphic and one antiholomorphic scalar, say $\Phi^1,\Phi^2,\overline{\Phi}_3$. 
In \cite{Mansson0703} it was shown that for any complex $q$  the 
Hamiltonian in this sector satisfies Reshetikhin's criteria for integrability. It is natural to 
wonder whether this case can also be seen to arise from our formalism, i.e. be understood at the level
of quantum symmetries of the classical Lagrangian. The immediate problem is that in this sector the
D--terms are not flavour--blind and thus contribute non--trivially to the Hamiltonian, with their 
contribution actually being crucial for integrability. 
A further  problem that arises when trying to apply our approach to that case
is related to that the XXZ spin chain arises there as a subspace. For
the XXZ spin chain Hamiltonian it is not possible to use the Hamiltonian as
a spectral--parameter--independent $\Rhat$ matrix 
(it gives too trivial a solution of the RTT relations). Even though for closed spin-chains
it is equivalent with the Hamiltonian consisting of Temperley-Lieb generators,
from the view of the Lagrangian they will always be distinguishable. 
To see the $U_q(su(2))$ quantum symmetry for the XXZ spin chain one needs the 
affine  symmetry. Using the spectral--parameter--dependent R--matrix one 
builds up a transfer matrix commuting with the Hamiltonian.

To make the above discussion more concrete we write out the scalar field
part of the Lagrangian which is responsible for the sector with 
$\phi^1$ and $\overline{\phi}^2$
\be
\widetilde{\phi}_{i}\widetilde{\phi}_{j}{H^{XXZ}}^{ij}_{kl}
\widetilde{\phi}^{k}\widetilde{\phi}^{l}, \qquad
\widetilde{\phi}_1=\overline{\phi}_1\,, \quad 
\widetilde{\phi}_2=\phi_2 \,,\quad
\widetilde{\phi}^1=\phi^1 \,,\quad
\widetilde{\phi}_1=\overline{\phi}^2
\ee 
with $H^{XXZ}$ being the nearest neighbour XXZ spin chain interaction. Up 
to a term proportional to the identity matrix, its nonzero 
elements can be normalised to (we just write $H$ from now on):
\be
H^{12}_{12}=H^{21}_{21}=Q \qquad H^{12}_{21}=H^{21}_{12}=-1 \;.
\ee
It is easy to see that we get more constraints from  the RTT relation
than we want for $\Rhat$.
The RTT relations for $\Rhat$ (or equivalently for $H$) give
\be
0=\tq 11 \tq 12 H^{12}_{12}+\tq 12 \tq 11 H^{21}_{12} \qquad
0=\tq 12 \tq 11 H^{21}_{21}+\tq 11 \tq 12 H^{12}_{21}\;.
\ee
The equation above leads to (when $Q\neq 1$ which is the non-deformed case)
\be
\tq 11 \tq 12=0 \qquad \mbox{and} \qquad \tq 12 \tq 11=0\;.
\ee
This clearly has more constraints that those coming from the Temperley-Lieb generator (\ref{Aq}).
 We hope to clarify how our construction extends beyond the holomorphic sector in future work.

\section{The classical $r$--matrix and noncommutativity} \label{Noncommutative}
 
 In this section we show how the quantum symmetries we have been discussing so far 
are linked to the previously known picture of the Leigh--Strassler marginal deformations as
\emph{non--commutative} deformations, in the sense of replacing standard multiplication by 
multiplication with a suitable star product. We begin with a short discussion of the classical
$r$--matrix, and show how this is related to the noncommutativity parameter appearing in the 
star product. We conclude with some comments on the dual gravity side.

\subsection{The classical $r$--matrix} \label{rmatrixsection}

Given the $R$--matrix (\ref{qRmatrix}) for the real $\beta$--deformation, we can take the
classical limit, which, for the (spectral--parameter--independent) case we are 
examining, corresponds to an expansion for small $\beta$. 
We thus write 
\be
\R i k j l=\delta^i_k \delta^j_l + i\beta r^{i\;j}_{k\;l}+O(\beta^2)
\ee
where $r$ is known as the classical $r$--matrix. Explicitly, for real $\beta$ it is given by

\be \label{classicalrrealbeta}
r=\text{diag}(0,-1,1,1,0,-1,-1,1,0)\;.
\ee
It satisfies the classical Yang--Baxter equation, 
\be \label{cYB}
[r_{12},r_{13}]+[r_{12},r_{23}]+[r_{13},r_{23}]=0\;.
\ee
The classical $r$--matrix as a limit of the $R$--matrix with spectral dependence, 
in the context of $\Ncal=4$ SYM has been discussed in \cite{Torrielli07,MoriyamaTorrielli07,BeisertSpill07,deLeeuw0804}, 
though of course there the classical limit was taken not with respect to the deformation parameter
$\beta$ (which was zero) but rather with respect to a suitable combination of momenta and the 
YM coupling constant (see e.g. \cite{BeisertSpill07} for further discussion of possible classical limits).

\subsection{Noncommutativity and Leigh--Strassler} \label{starproduct}

 It has been known for some time that the Leigh--Strassler theories are related to the introduction of 
non--commutativity in the geometry probed by the six scalars of $\Ncal=4$ SYM (thought of as the 
transverse coordinates to the stack of D3--branes used to define the theory), and therefore
to a star product between the fields. Focusing on the real $\beta$ case, Berenstein et al. 
\cite{Berensteinetal00} discussed the noncommutative structure 
of the moduli space of the theory parametrised by the vacuum expectation values of these scalars. On the amplitude side, 
\cite{KulaxiziZoubos04} introduced  a twistor--space star product for the full Leigh--Strassler deformation, which, 
however, was coordinate dependent and therefore not associative in general. 

Additional insight into the noncommutative structure of the Leigh--Strassler theories came with the
work of Lunin and Maldacena \cite{LuninMaldacena05}, who, as discussed above, constructed the AdS/CFT dual 
geometry of the real $\beta$ deformation. As part of their construction, these authors introduced a certain
star product between the scalar fields of $\Ncal=4$ SYM which concisely encoded the $\beta$ deformation.
Partly inspired by the work of \cite{LuninMaldacena05}, Gao and Wu \cite{GaoWu06} showed that for
real $\beta$ the twistor space star product of \cite{KulaxiziZoubos04} is indeed associative and 
one can thus use it to provide a consistent definition for (tree--level) amplitudes at all orders in the
deformation parameter. 
LM--type star products were also used in an essential way in \cite{Khoze05,Ananthetal06,Ozetal07}.
In \cite{Bundzik06}, the star--product approach was extended to the integrable cases (discussed in section
\ref{OtherIntegrable}) related to the $\beta$--deformation by changes of basis.

The main idea of \cite{LuninMaldacena05} was that, in the spirit of Seiberg and Witten \cite{SeibergWitten9908},  
noncommutativity on the gauge theory (open string side) would manifest itself as a deformed (but commuting) 
geometry plus NS and RR fields on the closed string side. 
Although this direct approach of mapping the noncommutativity parameter to a B--field was not the one that Lunin 
and Maldacena actually followed in constructing the dual background, it was later carried through in the 
work of \cite{Kulaxizi0610} (similar ideas were also discussed in \cite{CatalOzer05}, though without emphasising
the role of the noncommutativity parameter). 
What was shown in  \cite{Kulaxizi0610}  was that the $\beta$ deformation can be encoded by introducing the
following star--commutators between the coordinates of the $\Cset^3$ which is transverse to the stack of $D3$--branes:
\be \label{Starrelations}
[z^i,z^j]_*=i\beta \Theta^{ij}_{\;kl}z^k z^l,\quad
[z^i,\bar{z}^{\bar{j}}]_*=i\beta\Theta^{i\bar{j}}_{\;k\bar{l}}z^kz^{\bar{l}},\quad
[\bar{z}^{\bar{i}},\bar{z}^{\bar{j}}]_*=i\beta\Theta^{\bar{i}\bar{j}}_{\;\bar{k}\bar{l}}z^{\bar{k}}z^{\bar{l}}\;.
\ee
Here all effects of noncommutativity are encoded in the $*$--product, so in particular the $z$'s 
on the right--hand side are commuting. 
Note that there are non--trivial star products between holomorphic and antiholomorphic coordinates, 
which, translated to field theory language, turned out to be necessary in order for the D--terms 
to stay invariant under the noncommutative deformation \cite{Kulaxizi0610}.

Looking at the holomorphic sector, the explicit form of $\Theta^{ij}_{\; kl}$ in \cite{Kulaxizi0610} is 
\footnote{In \cite{Kulaxizi0610} $\Theta$ is given in the form $\Theta^{ij}=\Theta^{ij}_{\;kl}z^kz^l$. 
Since in that reference the $z$'s on the right--hand side were commuting coordinates 
(the noncommutativity being encoded in the star product) their ordering was not important, which
could potentially introduce an ambiguity between $\Theta^{ij}_{\;kl}$ and $\Theta^{ij}_{\;lk}$. 
However, the relation to the classical r--matrix naturally selects one of the orderings, namely
the one for which $\Theta^{ij}_{\;kl}$ is diagonal as a $9\times9$ matrix.} 
\be
\Theta=-\text{diag}(0,-1,1,1,0,-1,-1,1,0)\;,
\ee
i.e. the holomorphic noncommutativity parameter is simply the classical $r$--matrix (\ref{classicalrrealbeta}). 
To be precise, $r^{i\;j}_{k\;l}x^kx^l=-\Theta^{ij}$. 
Thus, in the holomorphic sector, the description of the
real $\beta$ deformation in \cite{Kulaxizi0610} (and therefore also in \cite{LuninMaldacena05}) via the 
introduction of non--commutativity between the scalars
of $\Ncal=4$ SYM is just a first--order manifestation of the quantum symmetry we have been discussing. 

However, in order to complete this identification, it is necessary to move beyond the holomorphic
sector. Can we make contact to the mixed and antiholomorphic relations in (\ref{Starrelations})?
In order to do so, it is more intuitive to go back to the geometry of the quantum plane. 

As we have seen, the holomorphic coordinates are taken to live on a quantum plane, defined by
\be
\R i k j l x^k x^l=x^jx^i\;.
\ee
At first order in $\beta$, these relations become 
\be \label{quantumplanefirstorderhol}
x^i x^j-x^j x^i=-i\beta r^{ij}_{\;kl}x^k x^l\;.
\ee
On translating from this quantum plane picture, where the coordinates are noncommuting, to an equivalent 
one where the coordinates commute, but noncommutativity is transferred to the star product, we immediately
see the equivalence  of (\ref{quantumplanefirstorderhol}) to the first of the star--product relations
in (\ref{Starrelations}).\footnote{For a deeper understanding of the relation between Hopf algebras
and noncommutativity than that provided here, see \cite{Watts99}.} 

As for the antiholomorphic coordinates, their commutation relations should clearly be obtained from
the definition of the co--plane (\ref{Ruu}):
\be
u_k u_l \R k i l j =u_j u_i\quad\longrightarrow\quad
 u_i u_j-u_j u_i=i\beta u_k u_l r^{kl}_{\;ij}\;.
\ee
This can also be seen to match the antiholomorphic star--product relation in (\ref{Starrelations}). 

Finally, we need to check the relations following from the mixed plane relations (\ref{MixedPlane}). 
First of all, expanding ${\tilde{R}}=1+i\beta\tilde{r}+O(\beta^2)$ it is easy to check that ${ \tilde{r}}=-{r}$ 
(this is true in general to first order in the parameters). 
So the relations in (\ref{MixedPlane}) reduce to:
\be
u_i x^j-x^j u_i=i\beta u_l r^{jl}_{\;ki}x^k, \;\quad x^i u_j-u_j x^i=-i\beta x^k r^{il}_{kj} u_l
\ee
which are of course consistent. Having these relations at hand, we can now straightforwardly check that 
the remaining $3\times 3$ blocks of the noncommutativity matrix are precisely reproduced. Therefore
we conclude that, for real $\beta$, our  understanding of the field theory deformations
as arising from the $R$--matrix (\ref{qRmatrix}) is consistent with the work of 
\cite{Kulaxizi0610} on the star product description of these theories. 

Of course the fact that the real $\beta$ story works so well is not that surprising. However,
 in \cite{Kulaxizi0612} this star--product approach was extended beyond the 
real $\beta$ case. The goal there was to attempt to construct the (still unknown) supergravity 
dual geometry of the general Leigh--Strassler theory by first understanding the open--string
side in terms of noncommutativity of the scalar fields (the transverse directions to the 
D3--brane defining the gauge theory) and then following the Seiberg--Witten procedure
to obtain the closed--string fields. The first step in this programme is thus to construct
the noncommutativity matrix $\Theta^{IJ}$ (c.f. (\ref{Starrelations})) describing the general 
deformation (here $I=\{i,\bar{i}\}, J=\{j,\bar{j}\}$). Initially restricting to the case of a purely 
$h$--deformation, and with the help of a number of assumptions (which are expanded on in 
\cite{Kulaxizi0612}) one arrives at a unique choice for $\Theta^{IJ}$ for $q=1, h=\hb=\rho_1$ as well as for
$q=1, h=-\hb=i\rho_2$. 
The classical $r$--matrices for these two cases are
\be
h=\rho_1: \quad r^{11}_{\;23}=r^{13}_{\;22}=-1\;, r^{11}_{\;32}=r^{12}_{\;33}=1 \;, \quad \text{and} \quad
h=i\rho_2: \quad r^{11}_{\;23}=r^{12}_{\;33}=1\;, r^{13}_{\;22}=r^{11}_{\;32}=-1 
\ee
plus cyclic permutations. Writing our holomorphic, antiholomorphic and mixed quantum plane relations 
for these two cases, we find that the resulting noncommutativity matrix precisely corresponds to
the $\Theta^{IJ}$ in 
\cite{Kulaxizi0612}. We believe that this lends support to that first step of the programme and to 
the assumptions used to derive the noncommutativity matrix. 

On the other hand, the next step, which involves using the Seiberg--Witten equations to derive
the closed--string background, is on less firm ground, since (as discussed in \cite{Kulaxizi0612})
the noncommutativity parameter does not seem to reduce to a constant matrix in an appropriate
coordinate system. So one is essentially applying the Seiberg--Witten equations beyond their
original regime of validity (of constant noncommutativity). A related problem is associated to 
the fact that this particular case is not integrable\footnote{This is in contrast to the $q=0,h\neq 0$ 
case in section \ref{qzerohnonzero}. However, as we discussed, that case does not seem to have a well--defined 
classical limit and might not be describable by a star product in a simple way.}. The immediate result of this is 
that the star product turns out not to be 
associative, which in \cite{Kulaxizi0612} led to a series of complications in constructing the
dual background. Despite this, a nontrivial solution of IIB supergravity was found in that work, 
up to third order in the deformation parameter. 

We do not know whether the methods of \cite{Kulaxizi0610,Kulaxizi0612} can 
be extended in order to construct the full supergravity solution. However, we believe
that, since our $R$--matrix approach places more emphasis on the symmetries of the problem, it could 
provide more insight on the underlying noncommutative geometry than that based purely on 
a star product and thus provide some useful input towards overcoming some of the problems
that arose there.

\section{Discussion and conclusions} \label{Conclusions}

In this work we identified and characterised the quantum symmetry (Hopf) algebra which 
underlies the Leigh--Strassler deformations of $\Ncal=4$ SYM. We did this by mapping the 
problem to that of understanding the symmetries of a particular type of cyclic quantum plane. 
The resulting algebra is a quantum deformation of the $\SU(3)$ R--symmetry present in the $\Ncal=4$ SYM
theory, and the commutation relations for the algebra generators are explicitly constructed from 
a generalised $R$--matrix via the standard RTT relations. However, this algebra is not one of the
standard multiparameter deformations of $\SU(3)$ known in the literature. 
In particular, since our $R$--matrix (\ref{qhRmatrix}) does not satisfy the 
Yang--Baxter equation for generic values of the parameters $q$ and $h$, the fact that there exists 
a solution to the RTT relations spanned by all the nine (before imposing the determinant condition)
generators of the algebra is nontrivial.

 A further complication arising from the violation of the Yang--Baxter equation 
is that the associativity condition for the Hopf algebra implies that the quadratic 
RTT relations generate new higher order (at least cubic) relations. Thus the algebra is not 
consistent as a quadratic algebra, but will rather be a higher order algebra.
Although this carries the danger of trivialising the algebra (by making the
ideal too large) we showed that this does not occur and that, in particular, 
the cubic relations are not in conflict with the existence of an antipode and
a central quantum determinant. 

Considering that quantum matrix Hopf algebras defined by generalised $R$--matrices (not
satisfying the YBE) have not received much attention in the literature, 
it would be of great interest to study them more and understand their consequences for 
various physical systems. A better understanding of our Hopf algebra would not only provide
more insight into the Leigh--Strassler deformation, but also into the spin--chain
Hamiltonian which the dilatation operator is mapped to. It would also be
interesting to understand its relation to the non--hermitian Hamiltonian
\cite{Mansson0703} obtained from the Belavin $R$--matrix \cite{Belavin81}, 
which generates the same quantum plane but not the same co--plane (since it is not
hermitian). Notice that for our proof that we have a Hopf algebra
we could treat $\qb$ and $\hb$ as independent of $q$ and $h$ and thus
it was not necessary for them to be complex conjugates of each other.
This means that included in the Hopf algebra we found is the case 
describing the Belavin non-hermitian Hamiltonian. It would be interesting 
to understand if this Hopf algebra is related to the elliptic quantum 
group which gives rise to the Belavin $R$--matrix.

Quantum symmetries have, for special values of $q$ and $h$, of course been observed before for the 
Leigh--Strassler theories in the context of spin chains, although they were 
never written down explicitly in the matrix quantum 
algebra picture, which we find the most intuitive one from a physicist's 
point of view (in the sense of being the most straightforward generalisation
of the usual matrix Lie algebra picture). The main novelty of our approach 
is that we can identify the quantum symmetry directly from the
(four--dimensional) field theory Lagrangian without needing to consider the corresponding
spin chain. From this point of view, the fact that the planar one--loop spin chain Hamiltonian 
enjoys this symmetry is simply due to the fact that the quantum (in the Hopf algebra sense) 
symmetry is not broken at the quantum (in the sense of planar gauge perturbation theory) level. This 
would also imply that this symmetry would remain beyond one--loop order, even though on the 
spin chain side we would have to consider long--range Hamiltonians\footnote{Another potential 
issue when considering the spin chain at higher loops is that the $\SU(3)$ sector ceases to be
closed, so one would need to consider at least the $\SU(2|3)$ sector (see \cite{Beisert05} for 
the $\Ncal=4$ discussion).}. Going beyond 
spin chains, one could hope that this quantum symmetry would provide some input towards algebraically 
determining the structure of the higher--loop finiteness conditions, thus (at least at the planar
level) potentially helping to characterise the parameter space of exactly marginal deformations, 
parametrised by the function $f(g,\kappa,q,h)=0$. We made some preliminary comments on this possibility in 
section \ref{qsandfiniteness}. 
 
 In the special cases where the generalised $R$--matrix reduces to an actual $R$--matrix satisfying
the Yang--Baxter equation, our Hopf algebra becomes an honest dual quasi--triangular Hopf algebra. This 
provides a very appealing explanation of why the generic LS theory is not (one--loop) integrable:
Integrability requires this quasi--triangular Hopf algebra structure which is not present in general.
 Thus we have come closer to the answer to the main questions posed
in the introduction: What is the crucial property that differentiates $\Ncal=4$ SYM, the real $\beta$
deformation and a few other examples of planar--integrable $4d$ field theories from the much
larger class of perturbatively finite $4d$ field theories? And, perhaps more importantly, is there a 
property that differentiates those latter cases from the far larger class of \emph{conformal} theories?
Of course, the fact that integrability requires an underlying quasi--triangular Hopf algebra structure is by
no means surprising. What we would like to advocate is that there might be a more general algebraic
structure underlying finiteness, some properties of which we have begun to uncover. Understanding
whether this is a general feature will require considering a wider range of finite theories beyond
the Leigh--Strassler deformations. 

 In this context, it is interesting to speculate whether there are other deformations of 
$\Ncal=4$ beyond the Leigh--Strassler examples which might be integrable. One direction to be
explored is clearly that of breaking supersymmetry by involving the $\Urm(1)_R$ factor in the
deformation \cite{Frolov05}. Thus one would be looking at integrable quantum deformations of $\SU(4)$. 
But another possibility would be to look at other deformations of $\SU(3)$, e.g. those appearing in the 
classification of \cite{EwenOgievetsky94}. Doing so would involve overcoming some immediate problems,
since, although these deformations do lead to $R$--matrices
satisfying the YBE, they also have several features which from a physical
point of view seem to make them undesirable in describing superpotential deformations. 
They generically break the cyclic $\Zset_3$  symmetry
(essential in the finiteness argument of \cite{LeighStrassler95}), and also do not seem
to lead to real Lagrangians (since $E_{ijk}$ and $F^{ijk}$ are not conjugates). 

 Eventually one would also like to explore how our algebra can be embedded into 
the full supergroup $\SU(2,2|4)$. Along these lines one
might try to make contact with the work of \cite{BeisertKoroteev08}, which
considered \emph{integrable} quantum deformations of the $\mathrm{psu}(2|2)\times\mathrm{I\!R}^3$
symmetry of the $\Ncal=4$ SYM S--matrix \cite{Beisert0511}. Since the
requirements of integrability and the $q$--deformed theories
do not match in general \cite{BerensteinCherkis04}, the connection of
our work to \cite{BeisertKoroteev08} is not immediately clear but certainly deserves
further study.

The fact that the associativity requirement on our Hopf algebra led to
cubic relations, and thus to the algebra not being consistent as a quadratic 
algebra, could perhaps be considered an unsatisfactory aspect of
our construction. An alternative approach would be to try to live 
with non--associativity. One well--known class of quantum algebras which 
have non--associativity built in (and under full control) 
is that of \emph{quasi}--Hopf algebras, introduced by Drinfel'd \cite{Drinfeld89}. In an
important subclass of these, called quasi--triangular quasi--Hopf, one 
considers a generalised $R$--matrix which satisfies a generalised 
Yang--Baxter equation. Furthermore, quasi--Hopf algebras have been 
considered in the past as suitable candidates for internal symmetries 
in field theory \cite{MackSchomerus92}. It would
certainly be very appealing if our algebra were to fall within this 
framework. We hope to report on this possibility in future work.  
 
 One fundamental element of any discussion of quantum groups and integrability 
which was conspicuous by its absence in this work is spectral--parameter dependence. 
Our general $R$--matrix (\ref{qhRmatrix}) clearly does not involve a spectral parameter. 
The spectral--parameter--dependent Hopf algebra structure in the $\Ncal=4$ context was essential in 
order to understand crossing and constrain the dressing phase \cite{Janik06,GomezHernandez06,Plefkaetal06}. 
Similarly, being able to introduce it here would certainly provide more insight into the 
structure of the Leigh--Strassler theories.  The standard procedure for introducing the spectral 
parameter is passing to the affine group, and it could be  worthwhile to understand whether it can be 
applied here. On the other hand, it is not clear whether we \emph{should} expect 
to see any spectral--parameter dependence at the level of the classical Lagrangian.

 With regard to the dual geometry, we should emphasise that one should probably not expect the 
quantum symmetry to be evident on the gravity side where, as observed in \cite{Berensteinetal00} in
a similar context, it is only the centre of the algebra which should be manifest. As discussed
in section \ref{starproduct}, one way to understand the effect of the quantum symmetry  on the dual geometry 
(at least to first order) is through open/closed duality and the ideas of \cite{SeibergWitten9908}. We should note,
however, that in applying the generalised geometry formalism to the real $\beta$ deformations 
\cite{Minasianetal06,HalmagyiTomasiello07,Granaetal08} one can actually identify the noncommutativity matrix on 
the gravity side as part of the construction. Perhaps this observation could be extended to the full quantum symmetry.  

Quantum groups are well--known to exhibit very interesting features at special values of the
parameters corresponding to roots of unity. In particular the representation structure at these
points is very different from the classical case. The Leigh--Strassler theories with $q$ a root
of unity have been studied in \cite{BerensteinLeigh00,Berensteinetal00}, both with regard to the
gauge theory moduli space as well as to their dual string backgrounds (which correspond to near--horizon 
limits of branes on orbifolds with discrete torsion). In order to make contact with and possibly extend 
that work, one would have to consider our algebra at roots of unity and understand the new features that 
might emerge there.

To conclude, using the Leigh--Strassler marginal deformations as our motivation, we have provided 
what we believe is a fresh and potentially unifying point of view on the interplay between 
integrability and finiteness in four--dimensional field theory. Although some aspects of our 
construction are perhaps tentative, we believe that it provides a useful starting point from
which to better understand the origins and consequences of integrability in field theory, as
well as a glimpse into what lies beyond.

\paragraph{Acknowledgments} 

 We would like to thank David Berenstein, Shinji Hirano, Peter Koroteev, 
Charlotte Kristjansen, Manuela Kulaxizi, Matthias Staudacher and G. Vartanov for 
very helpful discussions. K.Z. would like to thank the Albert Einstein 
Institute in Potsdam for its hospitality during the early stages of this 
work. He is also grateful to the organisers and
participants of the 6$^{\text{th}}$ Simons Workshop in Mathematics and 
Physics at Stony Brook for creating a stimulating scientific environment 
which aided the development of this project. T.M. would like to thank the
Niels Bohr Institute for its hospitality during the later stages of this work. 
She would also like to thank the Humboldt foundation for its support of 
financing this research. The research of
K.Z. is supported by Danish Research Council grant FNU-272-06-0434: 
``Spin Chains as the connecting link between String Theory and Gauge Theory''.

\appendix
\section{Definitions}

Since the language of quantum matrix bialgebras might not
be familiar to all readers, we will provide some of the most important definitions. 
For more details and proofs, the reader should consult one of the excellent references
on quantum groups and Hopf algebras, for instance \cite{Majid, Gomezetal,ChariPressley}. Much of the discussion 
below closely follows \cite{Majid}. 

\begin{figure}[ht]
\begin{center}
\begin{picture}(250,90)
\put(12,52){\vector(1,1){30}}
\put(82,52){\vector(-1,1){30}}
\put(41,6){\vector(-1,1){30}}
\put(54,6){\vector(1,1){30}}
\put(25,85){$\mathcal{C}\otimes \mathcal{C} \otimes \mathcal{C}$}
\put(75,40){$\mathcal{C}\otimes \mathcal{C}$}
\put(0,40){$\mathcal{C}\otimes \mathcal{C}$}
\put(44,0){$\mathcal{C}$}
\put(14,16){\small{$\Delta$}}
\put(70,16){\small{$\Delta$}}
\put(-7,64){\small{$\Delta \otimes Id$}}
\put(73,64){\small{$ Id \otimes \Delta$}}
\put(40,-17){(a)}
\put(202,60){\vector(-1,-2){20}}
\put(214,60){\vector(1,-2){20}}
\put(208,20){\vector(0,1){40}}
\put(196,64){$\mathcal{C}\otimes \mathcal{C}$}
\put(168,10){$k\otimes \mathcal{C}=\mathcal{C}=\mathcal{C}\otimes k$}
\put(163,36){\small{$\epsilon \otimes Id$}}
\put(229,36){\small{$Id \otimes\epsilon$}}
\put(199,36){\small{$\Delta$}}
\put(200,-17){(b)}
\end{picture}\\
\end{center}
\caption{A schematic representation of the action of the coproduct $\Delta$ and the counit $\epsilon$.} \label{coprod}
\end{figure}
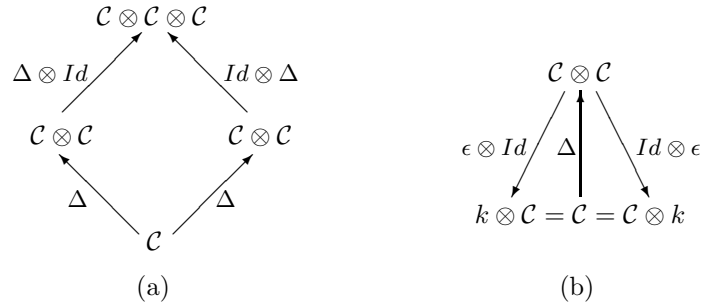

The main new features of bialgebras compared to algebras are the presence of a
\emph{coproduct} and a \emph{counit}, which act as shown in Fig. \ref{coprod}. 

 A \emph{coalgebra}
$(\mathcal{C},+, \Delta, \epsilon;k)$ over the field $k$ is a vector space $(\mathcal{C},+;k)$ over $k$ 
along with a linear coproduct map $\Delta$:  $\mathcal{C}\rightarrow \mathcal{C}\otimes \mathcal{C}$, 
which is coassociative and there exists a counit map $\epsilon: \mathcal{C} \rightarrow k$ as shown in the
figure. Note that reversing the arrows in the above figure gives an algebra.

A coalgebra is said to be \emph{cocommutative} if $\tau  \circ \Delta=\Delta $, where
$\tau$ is the transposition map: $\tau (v\otimes w)=w\otimes v$.

A \emph{bialgebra}  $(\Hcal,+,\cdot, \eta, \Delta, \epsilon;k)$
over $k$ (where $\cdot$ and $\eta$ are the standard algebra product and unit map respectively)
is a vector space $(\Hcal,+;k)$ which is both an algebra and a coalgebra in a compatible way:
\be
\Delta (hg)=\Delta (h)\Delta (g),
\qquad \Delta(1)=1\otimes 1\;, 
\qquad \epsilon (hg)=\epsilon (h)\epsilon (g)\;,  \qquad (h,g \in \Hcal)\;.
\ee
A \emph{Hopf algebra} is a bialgebra with an \emph{antipode} $S: \mathcal{C}\rightarrow \mathcal{C}$, which is an 
inverse--like object (though its action need not square to the unit element). The defining relations of the antipode are
\be
\cdot(S\otimes \mbox{id})\circ \Delta=\cdot(\mbox{id}\otimes S)\circ\Delta=\eta \circ\epsilon\;.
\ee
In the text we are interested in Hopf algebras which are quantum 
deformations of the algebra of functions of $\SU(3)$. However, in order to demonstrate the 
relevant concepts, let us briefly discuss the simpler case of $\GL(2)$ in some detail.
Let us start by considering an element $g$ of (the classical Lie group) $\GL(2)$. 
One way to describe the group is  through the  matrix entries $\{a,b,c,d\}$ of the group 
element ${\bf g}$ of $\GL(2)$
\be \label{GL2}
{\bf g}=\begin{pmatrix}a&b\cr c&d\cr \end{pmatrix}
\ee
with $ad-bc\neq 0$. The algebra, {\it fun}($\GL(2)$) of polynomial functions 
of the elements $\{a,b,c,d\}$ is a commutative algebra.
With the (non--cocommutative) coproduct defined as
\be \label{app:coproduct}
\Delta g^m_{\;n}=\sum_k g^m_{\;k}\otimes g^k_{\;n},
\qquad \epsilon g^m_{\;n}=\delta^m_{\;n}
\ee
(here $g^m_{\;n}$ denotes the elements of ${\bf g}$, for example $g^1_{\;1}=a$) and the antipode map
\be
S(g)\rightarrow g^{-1}
\ee
it becomes a commutative and non--cocommutative Hopf algebra.

A \emph{quasi--triangular Hopf algebra} is a Hopf algebra which is 
not cocommutative, but where the non--cocommutativity is controlled by a matrix $\Rcal$, called
the universal $R$--matrix. 
More precisely, it consists of a pair $(\Hcal,\Rcal)$, where $\Hcal$ is the Hopf algebra 
and ${\Rcal}$ is an invertible matrix in $\Hcal\otimes \Hcal$ and satisfies ($h \in \Hcal$)
\be
\label{AxiomsR}
\begin{split}
&(\Delta \otimes id)\circ \mathcal{R}=\mathcal{R}_{13}\mathcal{R}_{23},
\qquad (id \otimes \Delta)\circ \mathcal{R}=\mathcal{R}_{13}\mathcal{R}_{12}\\
& \tau \circ \Delta h =\Rcal h \Rcal^{-1}\;.
\end{split}
\ee
Using these axioms it is possible to show that $\Rcal$ 
satisfies an abstract Yang--Baxter equation. 

The existence of both a product and co-product leads to a natural notion of 
\emph{duality} for Hopf algebras: Given a Hopf algebra $\Hcal$ one can always define a 
dual Hopf algebra $\Hcal^*$ where the arrows in the diagrams above are interchanged. Thus
what is the coproduct in the original Hopf algebra becomes the product
in the dual Hopf algebra. This is the duality referred to in the text
when we mention that for some particular cases of the
parameters our Hopf algebra is dual to a quasi--triangular Hopf algebra.

All our discussion in the text is in the dual Hopf algebra picture, and in
particular the Hopf algebra $\Acal(R)$ defined by the relations in Table \ref{algebrarelations}
is really a dual Hopf algebra $\Hcal^*$ from the point of view of the definitions above. The reason
we take this perspective is clear from the definition (\ref{app:coproduct}) of the
coproduct (which we keep for $\Acal(R)$): It is what allows us to represent the algebra 
generators as matrices and work with them as we would in linear algebra. 
Notice that in the Hopf algebra dual to $\Acal(R)$ it is the non--cocommutativity which is 
controlled by an $R$--matrix, while in $\Acal(R)$ itself it is the noncommutativity 
which is controlled by an $R$--matrix. The fact that $\Acal(R)$ has the coproduct (\ref{app:coproduct})
is the reason that we can perform matrix multiplication of the elements ${\bf t}$
of $\Acal(R)$ as usual, while the fact that it is noncommutative is what causes the 
individual matrix components $\tq ij$ not to commute among themselves. If they did (i.e. $R$ 
became the unit matrix) we would reduce to the Lie algebra of $\SU(3)$ just as in the $\GL(2)$ 
example above. 

The duality between the two pictures is reflected by a difference in notation. Quantum groups
are usually defined in terms of a deformation of the universal enveloping algebra of a certain
undeformed Lie algebra, which e.g. for the standard $q$--deformation of $su(N)$ would be denoted 
as $U_q(su(N))$. The dual, linear algebra, picture would in this case be denoted by $\SU_q(N)$ 
(where the capital notation should not distract from the fact that quantum groups are Hopf
algebras and not groups). Although we could have chosen to employ the notation $\SU_{q,h}(3)$ to 
denote our two--parameter deformation of $\SU(3)$, this might allude to the standard deformations
of $\SU(3)$ which are different from ours, therefore we have simply denoted our matrix quantum
algebra $\Acal(R)$.

Let us now review some basic facts about the quasi--triangular Hopf algebra  $U_q(sl(2))$.
The defining commutation relations are \cite{Majid}:
\be
q^{\frac{H}{2}}X_{\pm}q^{-\frac{H}{2}}=q^{\pm}X_{\pm},
\qquad [X_+,X_-]=\frac{q^{H}-q^{-H}}{q-q^{-1}} \;.
\ee
This forms a Hopf algebra with coproduct
\be
\Delta q^{\pm \frac H2} =q^{\pm\frac{H}{2}}\otimes q^{\pm\frac{H}{2}},
\qquad \Delta X_{\pm}=X_{\pm}\otimes q^{\frac{H}{2}}+
q^{-\frac{H}{2}}\otimes X_{\pm}
\ee
(we suppress the explicit expressions for the counit and antipode).
The universal $R$--matrix $\Rcal$ related to the quasi--triangular structure is given by
\be \label{app:quasi--triangular}
\mathcal{R}=q^{\frac{H\otimes H}{2}}\sum_{n=0}^{\infty}
\frac{(1-q^{-2})^n}{[n]!}(q^{\frac{H}{2}}X_+\otimes q^{-\frac{H}{2}}X_-)^n
q^{\frac{n(n-1)}{2}}, \qquad \text{where} \quad [n]=\frac{q^n-q^{-n}}{q-q^{-1}} \;.
\ee
In order to exhibit this abstract $\Rcal$ as a concrete $R$--matrix we need to evaluate it
in a particular representation of the algebra. 
Let us choose the fundamental (spin--$\frac12$) representation as follows:
\be
\begin{split}
&H v_0=v_0, \qquad X_+v_0=0, \qquad X_- v_0=v_1 \\
&H v_1=-v_1, \qquad X_+v_1=v_0, \qquad X_- v_1=0
\end{split}
\ee
or in other words, choosing $v_0=\doublet 10$ and $v_1=\doublet 01$,  
\be
\rho(H)=\twobytwo100{-1}\;,\quad \rho(X_+)=\twobytwo 0100\;,\quad \rho(X_-)=\twobytwo 0010\;.
\ee
It is clear that out of the infinite series in (\ref{app:quasi--triangular}) only the first two terms are nonzero. 
We then find
\be
\begin{split}
R=(\rho\otimes\rho)(\mathcal{R})=&\twobytwo {q^{\frac14}}00{q^{-\frac14}}\otimes \twobytwo{q^{\frac14}}00{q^{-\frac14}}
\left(1\otimes1+(1-q^{-2}) \twobytwo 0{q^{\frac12}}00 \otimes \twobytwo 00{q^{\frac12}}0\right)\\
=&q^{-1/2}\left( \begin{array}{cccc} 
q&0&0&0\\0&1& q-q^{-1}&0\\0&0&1&0\\0&0&0&q\end{array} \right)\;.
\end{split}
\ee
Thus the matrix representation of the action of the universal $R$--matrix $\mathcal{R}$ is
just the $R$--matrix (\ref{R-matrix:suq2}).

\section{The quadratic bialgebra  relations} \label{quadraticappendix}
In this section we will show that any matrix $\hat{R}$ of the form
$\Rh ikjl=a\delta^i_{\;k}\delta^j_{\;l}+E_{klr}F^{rij}$ provides a non--trivial solution 
to the quadratic RTT equations (\ref{RTT}). Here the only restriction we put on $E$ and $F$ 
is that they are cyclic in the indices and that they are zero when two of the 
indices are alike but not the third. There is no loss in generality if we 
normalise the tensors $E_{ijk}$ and $F^{lmn}$ by setting $d=1$ in (\ref{qhchoice}), so we could choose to 
call the various non--zero elements
\be
\begin{split}
&E_{l(l+1)(l+2)}=1, \qquad E_{l(l+2)(l+1)}=-q, \qquad E_{lll}=h \\
&F^{l(l+1)(l+2)}=1, \qquad F^{l(l+2)(l+1)}=-\bar{q}, \qquad F^{lll}=\bar{h}\,.
\end{split}
\ee
In the general case we do not need to consider $(\bar{q},\bar{h})$ to be the complex
conjugates of $(q,h)$, so in all we do here we will consider
them to be linearly independent. Of course for the physical applications we consider
we will restrict to the special case where $\bar{q}$ and $\bar{h}$ are the complex 
conjugates of $q$ and $h$, as dictated by reality of the Leigh--Strassler Lagrangian.
Considering that the value of the constant $a$ does not affect the bialgebra 
we will choose it to be zero in the following.

 The RTT equations can be written in terms of $\hat{R}$ as
\be \label{RhTT}
\Rh a i b j \tq ic \tq jd = \tq ai \tq bj \Rh icjd 
\ee
or in terms of $E$ and $F$
\be
E_{ijl}F^{lab}\tq ic\tq jd =\tq ai \tq bj E_{cdl}F^{lij} \;.
\ee
On the left--hand side we  have three possibilities for $b$:
\be
b=a; \qquad b=a+1; \qquad b=a-1\,.
\ee
That means for a given $c$ and $d$ we have in total nine possibilities
on the left--hand side when we include the cyclic permutation of the indices
$a$. Equivalently
on the right--hand side we have three possibilities for $d$:
\be
d=c; \qquad d=c+1; \qquad d=c-1\,,
\ee
and all together for a given $a$ and $b$ we have in total nine possibilities
on the right--hand side. This gives in total 81 equations for our nine generators,
whose commutation relations we wish to know. If we require that all the
generators are non-trivial we should have just 36 (8+7+6+5+4+3+2+1)
commutation relations. In order for this to be consistent we need to show that 
the remaining equations are linearly dependent on these.

In order to keep track of all the equations we define the tensor 
\be
M^{ab}_{\; cd}:= E_{ijl}F^{lab}\tq ic\tq jd -\tq ai \tq bj E_{cdl}F^{lij} \;.
\ee
From $M^{a(a+1)}_{\; cc}=0$ we obtain
\be \label{Mrel1}
-q t^{a+1}_{\; c}\tq ac +\tq ac t^{a+1}_{\;c} =
h(t^a_{\; c+1} t^{a+1}_{\;c-1}
-t^a_{\; c-1} t^{a+1}_{\;c+1}\bar{q}+t^a_{\; c} t^{a+1}_{\;c}\bar{h})
-ht^{a-1}_{\;c}t^{a-1}_{\;c} \;.
\ee
Similarly from  $M^{a(a+1)}_{\; c(c-1)}=0$ we find
\be \label{Mrel2}
[t^{a+1}_{\; c},t^a_{\; c-1}] =
(-t^a_{\; c} t^{a+1}_{\;c-1}\bar{q}
+t^a_{\; c+1} t^{a+1}_{\;c+1}\bar{h})
+\frac{h}{q}t^{a-1}_{\;c}t^{a-1}_{\;c-1}
+\frac{1}{q}t^{a}_{\;c}t^{a+1}_{\;c-1} \;,
\ee
and $M^{a(a+1)}_{\; c(c+1)}=0$ gives
\be \label{Mrel3}
-qt^{a+1}_{\; c}t^a_{\; c+1}+\bar{q}t^{a}_{\; c+1}t^{a+1}_{\; c} =
t^a_{\; c-1} t^{a+1}_{\;c-1}\bar{h}
-h t^{a-1}_{\;c}t^{a-1}_{\;c+1} \;.
\ee
The above equations give us 27 equations. We can get nine more from
$M^{aa}_{\;cc}$:
\be \label{Mrel4}
t^{a}_{\; c+1}t^a_{\; c-1}-\bar{q} t^{a}_{\; c-1}t^{a}_{\; c+1} =
\frac{\bar{h}}{h}(t^{a+1}_{\; c} t^{a-1}_{\;c}
-q t^{a-1}_{\;c}t^{a+1}_{\;c})\;.
\ee
Thus the above gives us all the 36 commutators (or deformed commutators)
between the nine generators! These are tabulated in Table \ref{algebrarelations}.\footnote{Relation $(b)$
in Table \ref{algebrarelations} is actually a combination of (\ref{Mrel2}) and (\ref{Mrel3}) that we
found more useful in various manipulations.}

In order to prove our statement that there exist non--trivial solutions to (\ref{RhTT}) 
we now need to show that the remaining equations, encoded by $M^{(a+1)a}_{cc},M^{(a+1)a}_{c(c-1)},
M^{(a+1)a}_{(c-1)c},M^{cc}_{(a+1)a}$ and $M^{cc}_{a(a+1)}$
 are linearly dependent on the above.
Let us now show that. The following equations are all the same
\be
\begin{split}
&\frac{1}{h}\left( \bar{q}M^{a(a+1)}_{cc}+M^{(a+1)a}_{cc}\right)=
-\frac1{\bar{q}}\left(M^{a(a+1) }_{(c-1)(c+1)}+M^{(a+1)a}_{(c-1)(c+1)}\right)=
\bar{q}M^{a(a+1) }_{(c+1)(c-1)}+M^{(a+1)a}_{(c+1)(c-1)}= \\
& (\bar{q}^2 t^a_{\;c-1} t^{a+1}_{\;c+1}-t^{a+1}_{\; c+1}t^a_{\;c-1})
-\bar{q} (t^{a}_{\;c+1} t^{a+1}_{\;c-1}-
 t^{a+1}_{\;c-1} t^{a}_{\;c+1}) 
+\bar{h}(-\bar{q} t^{a}_{\;c} t^{a+1}_{\;c}-
 t^{a+1}_{\;c} t^{a}_{\;c}) \\
&=\frac1q\left(M^{a(a+1)}_{(c+1)(c-1)}-M^{(a-1)(a-1)}_{\;cc}-\bar{q}M^{a(a+1)}_{\;(c-1)(c+1)}\right)\;.
\end{split}
\ee
Note that the last expression contains only equations belonging to our original 36. 
The linear dependence of $M^{cc}_{(a+1)a}$ and $M^{cc}_{a(a+1)}$  follows from invariance under  exchanging
the upper and lower indices together with exchanging the barred parameters with the unbarred ones, from
which we obtain the equation
\be
 M_{(a+1) a}^{cc}+q M_{a(a+1)}^{cc}=
\frac{\bar{h}}{\bar{q}}\left(M_{a(a+1)}^{(c+1)(c-1)}-M_{(a-1)(a-1)}^{\;cc}-\bar{q}M_{a(a+1)}^{\;(c-1)(c+1)}\right)\;.
\ee
Finally, it is straightforward to show the following relation
\be
hM^{aa}_{c(c+1)}=M^{aa}_{(c-1)(c-1)}-\bar{h}M^{(a+1)(a-1)}_{(c-1)(c-1)}
+h\bar{h}M^{(a+1)(a-1)}_{c(c+1)}\;.
\ee
We have thus related all remaining $M^{ab}_{\;cd}$ to the four ones chosen above. We have therefore
proved that there exist non--trivial solutions to  the quadratic equations. The cubic relations following from 
them will be discussed in appendix \ref{cubicappendix}.

\section{The cubic bialgebra relations} \label{cubicappendix}

 There are two different ways to obtain cubic relations, either (c.f. (\ref{ch3:YB1}))
\be
\label{YB1}
\hat{R}_{12}\hat{R}_{23}\hat{R}_{12}{\bf t}_1{\bf t}_2 {\bf t}_3=
{\bf t}_1{\bf t}_2 {\bf t}_3\hat{R}_{12}\hat{R}_{23}\hat{R}_{12}
\ee 
or  (c.f. (\ref{ch3:YB2}))
\be
\label{YB2}
\hat{R}_{23}\hat{R}_{12}\hat{R}_{23}{\bf t}_1{\bf t}_2 {\bf t}_3=
{\bf t}_1{\bf t}_2 {\bf t}_3\hat{R}_{23}\hat{R}_{12}\hat{R}_{23} \;.
\ee
These two sets of equations are equivalent for any $R$--matrix in the
same equivalence class as one which satisfies the Yang-Baxter 
equation. For our generalised $R$--matrix this is not the case, and in the 
subsequent text we will analyse the extra cubic relations which follow
from these (for more discussion on this see section 3). As in appendix \ref{quadraticappendix},
it is sufficient to analyse the nontrivial $EF$ part of $\hat{R}$. 
To have the relations under better control we will again define some tensors
\be
\label{tensorYB}
\begin{split}
&M^{abc}_{def}:={M_L}^{abc}_{def}-{M_R}^{abc}_{def} \\
&N^{abc}_{def}:={N_L}^{abc}_{def}-{N_R}^{abc}_{def}
\end{split}
\ee
where
\be\label{MNdefinitions}
\begin{split}
&{M_L}^{abc}_{def}:=E_{ij\alpha}F^{\alpha ab}E_{kl\beta}F^{\beta jc}E_{mn\gamma}
F^{\gamma ik}\tq md\tq ne \tq lf \;,\\
&{M_R}^{abc}_{def}:=
\tq ai\tq bj \tq ck E_{lm \alpha}F^{\alpha ij}E_{nf\beta}F^{\beta mk}E_{de\gamma}
F^{\gamma ln} \;,\\
&{N_L}^{abc}_{def}:=E_{ij\alpha}F^{\alpha bc}E_{kl\beta}F^{\beta ai}E_{mn\gamma}
F^{\gamma lj}\tq kd\tq me \tq nf \;,\\
&{N_R}^{abc}_{def}:=
\tq ai\tq bj \tq ck E_{lm \alpha}F^{\alpha jk}E_{dn\beta}F^{\beta il}E_{ef\gamma}
F^{\gamma nm}\;.
\end{split}
\ee
In this new notation the equations (\ref{YB1}) and (\ref{YB2}) take the form
\be
\label{Cubicequations}
\begin{split}
&{M_L}^{abc}_{def}-{M_R}^{abc}_{def}=0 \\
&{N_L}^{abc}_{def}-{N_R}^{abc}_{def}=0\;.
\end{split}
\ee
Our main goal in this appendix is to show that these cubic relations are not in conflict
with the existence of an antipode and a central quantum determinant. The problem is that 
(\ref{Cubicequations}) are complicated relations which contain redundant information about the
algebra, so they are not immediately useful. We would like to manipulate them in order to find 
an irreducible set of cubic equations that are easier to work with. To do this,  
we will start by splitting them into two classes: The first one will be related to the diagonal components 
of the matrices ${\bf st}$ and ${\bf ts}$ (where ${\bf s}$ is the antipode (\ref{antipode})) while the second
class to the off--diagonal components. As we will see, the reason for treating them separately is
 that the first class does not lead to new cubic relations while the second one does. 

\mparagraph {The diagonal components}

 This is the case where both the upper indices of the $M_L$ and $N_L$ tensors
are either all equal or all different, and similarly for the lower set of indices of $M_R$ and $N_R$. 
 Let us write our cubic tensors a bit more explicitly by choosing a value for one index. This is
completely general, since the other choices can be recovered by cyclic symmetry. We obtain:
\be
\label{cubiceq1}
\begin{split}
{M_L}^{ab3}_{def} =
F^{ab3}&\left((\tq 1d \tq 2e \tq 3f-q\tq 2d\tq 1e\tq 3f+h\tq 3d\tq 3e\tq 3f)
(1+(\bar{q}q)^2+(\bar{h}h)^2)\right.\\
&\left.+(\tq 2d \tq 3e \tq 1f-q\tq 3d\tq 2e\tq 1f+h\tq 1d\tq 1e\tq 1f)
(\bar{q}q+\bar{h}h+\bar{q}q\bar{h}h) \right.\\
&\left.+(\tq 3d \tq 1e \tq 2f-q\tq 1d\tq 3e\tq 2f+h\tq 2d\tq 2e\tq 2f)
(\bar{q}q+\bar{h}h+\bar{q}q\bar{h}h)\right)\,,\qquad \{ a,b |\, F^{ab3}
\neq 0\}\,,
\end{split}
\ee
\be
\label{cubiceq2}
\begin{split}
{M_R}^{abc}_{de1} = 
 E_{de1}&\left((\tq a1 \tq b2 \tq c3-\bar{q}\tq a2\tq b1\tq c3+
\bar{h}\tq a3\tq b3\tq c3)
(\bar{q}q+\bar{h}h+\bar{q}q\bar{h}h)
\right. \\
&\left.+(\tq a3 \tq b1 \tq c2-\bar{q}\tq a1\tq b3\tq c2+
\bar{h}\tq a2\tq b2\tq c2)
(\bar{q}q+\bar{h}h+\bar{q}q\bar{h}h)\right. \\
& \left. +(\tq a2 \tq b3 \tq c1-\bar{q}\tq a3\tq b2\tq c1+\bar{h}\tq a1\tq b1\tq c1)
(1+(\bar{q}q)^2+(\bar{h}h)^2)\right),\qquad \{ d,e |\, E_{de1}\neq 0 \}\,,
\end{split}
\ee
\be
\label{cubiceq3}
\begin{split}
{N_L}^{3bc}_{def} =
F^{3bc}&\left((\tq 3d \tq 1e \tq 2f-q\tq 3d\tq 2e\tq 1f+h\tq 3d\tq 3e\tq 3f)
(1+(\bar{q}q)^2+(\bar{h}h)^2)\right.\\
&\left.+(\tq 1d \tq 2e \tq 3f-q\tq 1d\tq 3e\tq 2f+h\tq 1d\tq 1e\tq 1f)
(\bar{q}q+\bar{h}h+\bar{q}q\bar{h}h) \right.\\
&\left.+(\tq 2d \tq 3e \tq 1f-q\tq 2d\tq 1e\tq 3f+h\tq 2d\tq 2e\tq 2f)
(\bar{q}q+\bar{h}h+\bar{q}q\bar{h}h)\right)\,,
\qquad \{ b,c |\, F^{3bc}\neq 0\}
\end{split}
\ee
and
\be
\label{cubiceq4}
\begin{split}
{N_R}^{abc}_{1ef} = 
 E_{1ef}&\left((\tq a3 \tq b1 \tq c2-\qb\tq a3\tq b2\tq c1+\hb\tq a3\tq b3\tq c3)
(\bar{q}q+\bar{h}h+\bar{q}q\bar{h}h)
\right. \\
&\left.+(\tq a2 \tq b3 \tq c1-\qb\tq a2\tq b1\tq c3+\hb\tq a2\tq b2\tq c2)
(\bar{q}q+\bar{h}h+\bar{q}q\bar{h}h)\right. \\
& \left. +(\tq a1 \tq b2 \tq c3-\qb\tq a1\tq b3\tq c2+\hb\tq a1\tq b1\tq c1)
(1+(\bar{q}q)^2+(\bar{h}h)^2)\right),
\qquad \{ e,f |\,E_{1ef}\neq 0\}\;.
\end{split}
\ee
First we will prove that the equations (\ref{cubiceq1}) and
(\ref{cubiceq2}) do not lead to any cubic relations. Then we will use
the fact that $N$ and $M$ are related through the combined operation
of switching the upper and lower indices, and switching barred and unbarred parameters. 
To do this we will find it useful to consider special
linear combinations of the tensors which considerably simplify the calculation.
Before writing them down, we will make the following definitions
\be
\begin{split}
&x:=\left(\qb(\tq 11 \tq 22 \tq 33-\bar{q}\tq 12\tq 21\tq 33+
\bar{h}\tq 13\tq 23\tq 33)
+(\tq 21 \tq 12 \tq 33-\bar{q}\tq 22\tq 11\tq 33+
\bar{h}\tq 23\tq 13\tq 33)\right)\;,\\
&y:=\left(\qb(\tq 12 \tq 23 \tq 31-\bar{q}\tq 13\tq 22\tq 31+
\bar{h}\tq 11\tq 21\tq 31)
+(\tq 22 \tq 13 \tq 31-q\tq 23\tq 12\tq 31+h\tq 21\tq 11\tq 31)\right)\;,\\
&z:=\left(\qb(\tq 13 \tq 21 \tq 32-\bar{q}\tq 11\tq 23\tq 32+
\bar{h}\tq 12\tq 22\tq 32)
+(\tq 23 \tq 11 \tq 32-\bar{q}\tq 21\tq 13\tq 32+
\bar{h}\tq 22\tq 12\tq 32)\right)\;,\\
&A:=1+(\bar{q}q)^2+(\bar{h}h)^2\;, \quad B:=\bar{q}q+\bar{h}h+\bar{q}q\bar{h}h\;.
\end{split}
\ee
Now we write down some linear combinations of the $M$ tensors that lead
to nice looking equations:
\be
\begin{split} \label{formshort}
&\qb M^{123}_{de3}+M^{213}_{de3}=0 \qquad \Rightarrow \qquad
 \qb {M_R}^{123}_{de3}+{M_R}^{213}_{de3}=0 \qquad\Rightarrow \qquad
Ax+By+Bz=0 \\
&\qb M^{123}_{de1}+M^{213}_{de1}=0 \qquad \Rightarrow \qquad
\qb {M_R}^{123}_{de1}+{M_R}^{213}_{de1}=0
 \qquad \Rightarrow \qquad Bx+Ay+Bz=0 \\
&\qb {M}^{123}_{de2}+{M}^{213}_{de2}=0  \qquad \Rightarrow \qquad
\qb {M_R}^{123}_{de2}+{M_R}^{213}_{de2}=0
\qquad \Rightarrow \qquad Bx+By+Az=0 \;.
\end{split}
\ee
Note that the first step (canceling the $M_L$ parts) is possible for any values of the $e,f$ indices, 
however for the second step we need to require the condition in (\ref{cubiceq2}). 
The three equations in (\ref{formshort})  imply that each of $x,y$ and $z$ must be zero 
(unless $A=B$) and the resulting cubic conditions all have the form e.g.:  
\be
\label{Cubicrelation2}
\qb\left(\tq 11 \tq 22 -\qb\tq 12\tq 21+\hb\tq 13\tq 23
-\qb^{-1}(\tq 12 \tq 21 -\qb\tq 22\tq 11+\hb\tq 32\tq 31)\right)\tq 33=0 \;.
\ee
However, it follows from the quadratic relations that the expression inside  
the parentheses is zero, and thus for this choice of indices in the
tensor $M$ the quadratic relations did not induce further cubic 
restrictions. The same happens for $N$ starting from (\ref{cubiceq3}) and (\ref{cubiceq4}). 
Thus, if these were all the possibilities for
the indices, the story would have been over and we would have had a 
consistent quadratic algebra. But now this is not the end of a story, but
the beginning of a new one.

\mparagraph{The off--diagonal components}

Now we consider the cases where the upper indices of $M$ and $N$ are such that 
two are equal and the third is different. In this case we will see that
the quadratic relations lead us to cubic ones as a consequence of the different 
possible orderings. Consider e.g. ${M_L}^{112}_{def}$:
\be
\begin{split}
{M_L}^{112}_{def} &=\bar{h}\left(
(\tq 1d \tq 2e \tq 1f-q\tq 2d\tq 1e\tq 1f+h\tq 3d\tq 3e\tq 1f)
(\bar{q}^2h-q\bar{h}^2-qh)\right.\\
&\left.+(\tq 3d \tq 1e \tq 3f-q\tq 1d\tq 3e\tq 3f+h\tq 2d\tq 2e\tq 3f)
(q^2\bar{h}-\bar{q}h^2-\bar{q}\bar{h})\right.\\
&\left.+(\tq 2d \tq 3e \tq 2f-q\tq 3d\tq 2e\tq 2f+h\tq 1d\tq 1e\tq 2f)
(\bar{q}q+\bar{h}h+\bar{q}q\bar{h}h)\right)\;.\\
\end{split}
\ee 
From this, and the similar expression for ${M_L}^{122}_{def}$, we deduce the following relations
\be
\label{Mrelations}
-\frac1{\bar{q}}{M_L}^{322}_{def} ={M_L}^{232}_{def}=
\frac1{\bar{h}}{M_L}^{112}_{def} \quad\text{and}\quad  {M_L}^{122}_{def}=
-\frac1{\bar{q}}{M_L}^{212}_{def}=\frac1{\bar{h}}{M_L}^{332}_{def}\;.
\ee 
Similarly for  $N_L$ we find
\be
\begin{split}
{N_L}^{112}_{def} &=\left(
(\tq 3d \tq 3e \tq 1f-q\tq 3d\tq 1e\tq 3f+h\tq 3d\tq 2e\tq 2f)
(\bar{q}^2h-q\bar{h}^2-qh)\right.\\
&\left.+(\tq 2d \tq 2e \tq 3f-q\tq 2d\tq 3e\tq 2f+h\tq 2d\tq 1e\tq 1f)
(q^2\bar{h}-\bar{q}h^2-\bar{q}\bar{h})\right.\\
&\left.+(\tq 1d \tq 1e \tq 2f-q\tq 1d\tq 2e\tq 1f+h\tq 1d\tq 3e\tq 3f)
(\bar{q}q+\bar{h}h+\bar{q}q\bar{h}h)\right)\;.\\
\end{split}
\ee 
Considering also ${N_L}^{122}_{def}$, we obtain
\be
\label{Nrelations}
-\frac1{\bar{q}}{N_L}^{121}_{def}={N_L}^{112}_{def}=
\frac1{\bar{h}}{N_L}^{133}_{def} \quad \text{and}\quad 
{N_L}^{131}_{def} =-\frac1{\bar{q}}{N_L}^{113}_{def}=
\frac1{\bar{h}}{N_L}^{122}_{def}\;.
\ee
The above equations (\ref{Mrelations}) and (\ref{Nrelations}) will guide us to 
again make appropriate choices of linear combinations of the $M$ and $N$ in order
to reach irreducible cubic relations. Consider:
\be
\begin{split}
& \qb M^{112}_{123}+\bar{h} M^{322}_{123}=0 \quad\Rightarrow \quad 
 \qb {M_R}^{112}_{123}+\bar{h} {M_R}^{322}_{123}=0 \quad \Rightarrow \\
&\left(\qb(\tq 11 \tq 12 \tq 23-\qb\tq 12\tq 11\tq 23+\hb\tq 13\tq 13\tq 23)
+\bar{h}(\tq 31 \tq 22 \tq 23-\qb\tq 32\tq 21\tq 23+
\hb\tq 33\tq 23\tq 23)\right)
(1+(\bar{q}q)^2+(\bar{h}h)^2)\\
&+\left(\qb(\tq 12 \tq 13 \tq 21-\qb\tq 13\tq 12\tq 21+\hb\tq 11\tq 11\tq 21)
+\bar{h}(\tq 32 \tq 23 \tq 21-\qb\tq 33\tq 22\tq 21+
\hb\tq 31\tq 21\tq 21)\right)
(\bar{q}q+\bar{h}h+\bar{q}q\bar{h}h)\\
&+\left(\qb(\tq 13 \tq 11 \tq 22-\qb\tq 11\tq 13\tq 22+\hb\tq 12\tq 12\tq 22)
+\bar{h}(\tq 33 \tq 21 \tq 22-\qb\tq 31\tq 23\tq 22+
\hb\tq 32\tq 22\tq 22)\right)
(\bar{q}q+\bar{h}h+\bar{q}q\bar{h}h)=0\;.
\end{split}
\ee
Now by cyclically permuting the lower indices we get all in all three equations
again of the form (\ref{formshort}) (but now with the $x$, $y$ and $z$
instead being what we have in the parentheses above). This leads to  
\be
\label{Cubicrelation3a}
\qb(\tq 11 \tq 12 \tq 23-\qb\tq 12\tq 11\tq 23+\hb\tq 13\tq 13\tq 23)
+\bar{h}(\tq 31 \tq 22 \tq 23-\qb\tq 32\tq 21\tq 23+
\hb\tq 33\tq 23\tq 23)=0\;,
\ee
\be
\label{Cubicrelation3b}
\qb(\tq 12 \tq 13 \tq 21-\qb\tq 13\tq 12\tq 21+\hb\tq 11\tq 11\tq 21)
+\bar{h}(\tq 32 \tq 23 \tq 21-\qb\tq 33\tq 22\tq 21+
\hb\tq 31\tq 21\tq 21)=0\;,
\ee
\be
\label{Cubicrelation3c}
\qb(\tq 13 \tq 11 \tq 22-\qb\tq 11\tq 13\tq 22+\hb\tq 12\tq 12\tq 22)
+\bar{h}(\tq 33 \tq 21 \tq 22-\qb\tq 31\tq 23\tq 22+\hb\tq 32\tq 22\tq 22)=0\;.
\ee
Now we introduce the following useful tensor
\be
L^{abc}=(\tq a1 \tq b2 \tq c3\!-\!\qb\tq a2\tq b1\tq c3\!+\!\hb\tq a3\tq b3\tq c3)
\!+\!(\tq a2 \tq b3 \tq c1\!-\!\qb\tq a3\tq b2\tq c1\!+\!\hb\tq a1\tq b1\tq c1)
\!+\!(\tq a3 \tq b1 \tq c2\!-\!\qb\tq a1\tq b3\tq c2\!+\!\hb\tq a2\tq b2\tq c2).
\ee
In this notation we can write the sum of the equations above as
\be
\label{finalcubic1}
(\ref{Cubicrelation3a})+(\ref{Cubicrelation3b})+(\ref{Cubicrelation3c})=
\qb L^{112}+\hb L^{322}=0
\ee
and we also have all the cyclic permutations of the equation above.

Then performing the same manipulations with the tensor $N$ as we just did for $M$ we find
\be
\begin{split}
&\bar{h} N^{112}_{312}- N^{133}_{312}=0 \Rightarrow 
\bar{h} {N_R}^{112}_{312}-{N_R}^{133}_{312}=0 \Rightarrow \\
&\left(\bar{h}(\tq 13 \tq 11 \tq 22-\qb\tq 13\tq 12\tq 21+\hb\tq 13\tq 13\tq 23)
-(\tq 13 \tq 31 \tq 32-\qb\tq 13\tq 32\tq 31+\hb\tq 13\tq 33\tq 33)\right)
(1+(\bar{q}q)^2+(\bar{h}h)^2)\\
&+\left(\bar{h}(\tq 11 \tq 12 \tq 23-\qb\tq 11\tq 13\tq 22+
\hb\tq 11\tq 11\tq 21)
-(\tq 11 \tq 32 \tq 33-\qb\tq 11\tq 33\tq 32+\hb\tq 11\tq 31\tq 31)\right)
(\bar{q}q+\bar{h}h+\bar{q}q\bar{h}h)\\
&+\left(\bar{h}(\tq 12 \tq 13 \tq 21-\qb\tq 12\tq 11\tq 23+\hb\tq 12\tq 12\tq 22)
-(\tq 12 \tq 33 \tq 31-\qb\tq 12\tq 31\tq 33+\hb\tq 12\tq 32\tq 32)\right)
(\bar{q}q+\bar{h}h+\bar{q}q\bar{h}h)=0
\end{split}
\ee
plus using once again that the equation above plus the ones we get
from cyclically permuting are of the form (\ref{formshort}) we have:
\be
\label{Cubicrelation4a}
\bar{h}(\tq 13 \tq 11 \tq 22-\qb\tq 13\tq 12\tq 21+\hb\tq 13\tq 13\tq 23)
-(\tq 13 \tq 31 \tq 32-\qb\tq 13\tq 32\tq 31+\hb\tq 13\tq 33\tq 33)=0\;,
\ee
\be
\label{Cubicrelation4b}
\bar{h}(\tq 11 \tq 12 \tq 23-\qb\tq 11\tq 13\tq 22+
\hb\tq 11\tq 11\tq 21)
-(\tq 11 \tq 32 \tq 33-\qb\tq 11\tq 33\tq 32+\hb\tq 11\tq 31\tq 31)=0\;,
\ee
\be
\label{Cubicrelation4c}
\bar{h}(\tq 12 \tq 13 \tq 21-\qb\tq 12\tq 11\tq 23+\hb\tq 12\tq 12\tq 22)
-(\tq 12 \tq 33 \tq 31-\qb\tq 12\tq 31\tq 33+\hb\tq 12\tq 32\tq 32)=0\;.
\ee
Again the sum of the above equations can be expressed in terms of $L$
\be
\label{finalcubic2}
(\ref{Cubicrelation4a})+(\ref{Cubicrelation4b})+(\ref{Cubicrelation4c})=\hb L^{112} -L^{133}=0
\ee
plus cyclic permutations of this.
Now we can use equations (\ref{finalcubic1}) and (\ref{finalcubic2}) in alternation
(together with their cyclically permuted versions ) to show the following
\be \label{alternation}
\qb L^{112}=-\hb L^{322}=-\hb^2 L^{331}=\frac{\hb^3}{\qb} L^{211}=\frac{\hb^4}{\qb} L^{223}= 
-\frac{\hb^5}{\qb^2} L^{133}=-\frac{\hb^6}{\qb^2} L^{112} \;.
\ee
Thus we see that, for generic values of the parameters, $L^{112}$ is zero. Writing it out explicitly,
\be \label{L112}
(\tq 13 \tq 11 -\qb\tq 11\tq 13+\hb\tq 12\tq 12)\tq 22
+(\tq 11 \tq 12 -\qb\tq 12\tq 11+\hb\tq 13\tq 13)\tq 23
+(\tq 12 \tq 13 -\qb\tq 13\tq 12+\hb\tq 11\tq 11)\tq 21=0\;.
\ee
The quadratic expressions inside the parentheses are non--zero, thus
the quadratic relations have generated cubic relations. 
As a consequence of this the ideal generated by the quadratic relations includes
cubic relations. These relations are the ones responsible for the vanishing of
off--diagonal terms in (\ref{ch4:offdiagonal2}), where we prove the existence of the antipode.

 As a consistency check, it can easily be verified that if we restrict to the real $\beta$ deformation 
(where $\qb=1/q, h=0$) this cubic relation does follow from the quadratic ones in (\ref{Qalgebraq}). 
This is as it should be because that case is dual quasi--triangular and there should be no new equations at
cubic order. 

Clearly all the other tensors of type $L^{a(a\!-\!1)(a\!-\!1)}$ and $L^{aa(a\!+\!1)}$ in (\ref{alternation}) vanish.
Similar manipulations show that the $L^{aa(a\!-\!1)}$ and $L^{a(a\!+\!1)(a\!+\!1)}$ tensors vanish as well. For 
clarity we write $L^{122}$ explicitly:
\be \label{L122}
\tq 11(\tq 22 \tq 23 -\qb\tq 23\tq 22+\hb\tq 21\tq 21)
+\tq 13(\tq 21 \tq 22 -\qb\tq 22\tq 21+\hb\tq 23\tq 23)
+\tq 12(\tq 23 \tq 21 -\qb\tq 21\tq 23+\hb\tq 22\tq 22)=0 \,.
\ee
Finally, from equation (\ref{MNdefinitions}) it is clear that 
$M_R$ and $M_L$ and respectively $N_R$ and $N_L$ are related to each other
through exchanging the upper indices with the lower indices at the same time as
exchanging the $E_{ijk}$ with $F^{ijk}$. This implies that from knowing that 
we have the constraint (\ref{L122}) we also know that we will have constraints 
of the form:
\be \label{L122conj}
\tq 11(\tq 22 \tq 32 -q\tq 32\tq 22+h\tq 12\tq 12)
+\tq 31(\tq 12 \tq 22 -q\tq 22\tq 12+h\tq 32\tq 32)
+\tq 21(\tq 32 \tq 12 -q\tq 12\tq 32+h\tq 22\tq 22)=0 \,.
\ee 
It is possible that many of the resulting equations will be linearly dependent, or it could be that put together 
they impose even stronger cubic constraints on the algebra. We have not performed a thorough analysis of these constraints,
so we do not claim to have found an irreducible set of equations. For our present purposes (proving the existence of the
antipode and central determinant) it is enough that relations like (\ref{L112}) and (\ref{L122conj}) exist and that
they can never interfere with the cubic relations that need to stay nonzero. In particular, they do not 
force ${\mathbb D}=0$. There will also be further equations arising from the sector where both the upper and 
lower sets of indices contain two equal indices. These relations are also not relevant for our present purposes,
since they do not arise in the proof of the existence of the antipode. We hope to report on a full analysis of 
the cubic relations, as well as the possibility of higher--order relations, in a future publication.

\section{On matrix representations of the algebra generators} \label{matrixreps}

In this appendix we discuss the possibility of finding explicit matrix representations
of our algebra generators $\tq ij$, satisfying the relations in Table \ref{algebrarelations} which define
our Hopf algebra $\Acal(R)$. Recall that these were found through the FRT relations
from the R-matrix (\ref{qhRmatrix}), after taking into account the symmetries and
simplifying them to obtain an independent set of quadratic relations.

Using these relations, we then showed that transforming the chiral superfields 
$\Phi^i$ as $\Phi^i\ra \tq ij \Phi^j$
leaves the superpotential (4.5) invariant. 
Here of course the $\tq ij$ depend neither on spacetime nor on the 
fields $\Phi^i$, since the symmetry $\Acal(R)$ is just a deformation of the global  $\SU(3)$
symmetry group of
the undeformed case. Note that in the undeformed case $q\ra 1$, all the  $\tq ij$ commute and can be 
taken to be numbers, and thus the matrix 
\be
\mathrm{t}= \begin{pmatrix} \tq 11 &\tq 12 &\tq 13\\
\tq 21 &\tq 22&\tq 23 \\ \tq 31 &\tq 32 &\tq 33
\end{pmatrix}
\ee
is simply an $\SU(3)$ matrix. But, for general $q$, the elements $\tq ij$ are not numbers, but operators. 

Although the definition of the operators $\tq ij$ through their commutation relations (which
we showed to be consistent, i.e. they lead to a non-trivial algebra) is 
sufficient for our purposes, it might be interesting to check whether 
it is possible to represent them as matrices in some auxiliary space. 
We would thus be looking for 9 matrices $\rho(\tq 11),\rho(\tq 12),\cdots, \rho(\tq 33)$ 
satisfying the 36 
relations in Table \ref{algebrarelations}. 
It does not seem likely that they can be represented in the space of 
\emph{finite}-dimensional matrices in the general case. But it is certainly possible 
if one restricts to a (closed) subset 
of the $\tq ij$. This is well known in the quantum group literature, where,  
for example, for the standard $\SU_q(3)$ (or $\SL_q(3)$ in case of complex $q$) $R$-matrix, one can define the representation \cite{Majid}
\be
\rho(\tq ab)^i_{\;j}=R^{a\; i}_{\;b\;j}
\ee
which explicitly gives 
\be
\begin{split}
\rho(\tq 11)&=\left( \begin{array}{ccc} 
q&0&0\\0&1& 0\\0&0&1\end{array} \right),\qquad 
\rho(\tq 22)=\left( \begin{array}{ccc} 
1&0&0\\0&q& 0\\0&0&1\end{array} \right),\qquad 
\rho(\tq 33)=\left( \begin{array}{ccc} 
1&0&0\\0&1& 0\\0&0&q\end{array} \right), \\
\rho(\tq12)&=\left( \begin{array}{ccc} 
0&0&0\\q-q^{-1}&0& 0\\0&0&0\end{array} \right),\quad 
\rho(\tq23)=\left( \begin{array}{ccc} 
0&0&0\\0&0& 0\\0&q-q^{-1}&0\end{array} \right),\quad 
\rho(\tq13)=\left( \begin{array}{ccc} 
0&0&0\\0&0& 0\\q-q^{-1}&0&0\end{array} \right).
\end{split}
\ee
with the rest of the $\rho(\tq ij)$ being zero. These satisfy the standard $\SU_q(3)$ commutation relations.

Our $R$-matrix is not the standard $\SU_q(3)$ one, but we can still try to look for 
similar subsets of the $\tq ij$. An interesting subset is the one obtained 
by restricting to the (closed) subgroup given by $\tq 11,\tq 12,\tq 21,\tq 22$ and $\tq33$, i.e.
\be
\mathrm t=\begin{pmatrix}\tq 11 &\tq 12&0\\ \tq 21 &\tq 22 &0 \\0&0& 1\!\!1\end{pmatrix}
\ee
where we have chosen $\tq 33=1\!\!1$ (the unit matrix in the auxiliary space) and all the rest to be zero. 
Further taking $h=\bar{h}=0$, and $q$ real, we find that the commutation relations in Table \ref{algebrarelations} reduce to the following:
\be \label{commrel}
\begin{split}
\tq 11 \tq 12&=q \tq 12\tq 11\;,\quad \tq 12\tq 22=q \tq 22\tq12\;,\quad
[\tq 11,\tq 22]=q\tq 21\tq 12-q^{-1}\tq 12\tq 21\;,\quad\\
\tq21&\tq12=\tq 12\tq 21\;,\quad
\tq 11 \tq 21=q \tq21 \tq 11\;,\quad \tq 21\tq 22=q \tq 22\tq21\;,
\end{split}
\ee
which, after redefining $q\ra q^{-1}$, are precisely the commutation relations of standard $\SU_q(2)$ 
((\ref{Aq}) for $q$ real). Thus our commutation relations contain standard $\SU_q(2)$ 
as a special case.\footnote{The representations of this quantum group
are most conveniently studied in terms of the dual $U_q(su(2))$ algebra \cite{Majid}.}

For the case $h=0$ and $q$ real we can also consider the following (slightly degenerate) 
subset given by only $\tq 11,\tq 12,\tq22$ and $\tq33$.
In this simple case the commutation relations become:
\be
\tq 11\tq 12=q\tq21\tq11\;,\quad \tq 22\tq12=q^{-1}\tq 12\tq 22\quad\text{and}\;\;\tq11\tq22=\tq22\tq11
\ee
and it is trivial to find explicit matrix representations. One possibility is  
in terms of a 3-dimensional auxiliary space (from now on we write just $\tq ij$ instead of $\rho(\tq ij)$):
\be
\tq 11=a\cdot\begin{pmatrix} q^{-1} & 0&0\\ 0&1&0\\0&0&1\end{pmatrix},\;\;
\tq 22=b\cdot\begin{pmatrix} q & 0&0\\ 0&1&0\\0&0&1\end{pmatrix},\;\;
\tq 33=c\cdot\begin{pmatrix} 1 & 0&0\\ 0&1&0\\0&0&1\end{pmatrix},\;\;
\tq 12=d\cdot\begin{pmatrix} 0 & 0&0\\ 1&0&0\\0&0&0\end{pmatrix},\;\;
\ee
with $a,b,c,d$ complex numbers. Note that $a\cdot b\cdot c=1$ since the quantum determinant in this 
case is simply $\det_q\mathrm t=\tq 11\tq22\tq33=1$. For $q\ra 1$ all matrices commute. 

For clarity, let us write the transformation of the superpotential explicitly in this special case:
\be
\begin{split}
&\Phi^1\Phi^2\Phi^3-q\Phi^2\Phi^1\Phi^3\rightarrow
(\tq 11 \Phi^1+\tq 12 \Phi^2)\tq 22\Phi^2\tq 33\Phi^3-q\tq 22\Phi^2(\tq 11\Phi^1+\tq12\Phi^2)\tq33\Phi^3\\
&=\tq 11\tq 22\tq 33 \Phi^1\Phi^2\Phi^3+\tq 12\tq 22\tq 33 \Phi^2\Phi^2\Phi^3
-q\tq22\tq11\tq33\Phi^2\Phi^1\Phi^3-q\tq22\tq12\tq33\Phi^2\Phi^2\Phi^3
\end{split}
\ee
where we have used that (as always for non-braided quantum groups) 
the $\tq ij$ commute with the quantum plane elements $\Phi^i$.
From here we can either proceed using the commutation relations (\ref{commrel}), or, 
since we have a matrix representation, by converting the $\tq ij$ to matrices in the auxiliary space:
\be
\begin{split}
&\tq 11\tq 22\tq 33 \Phi^1\Phi^2\Phi^3+\tq 12\tq 22\tq 33 \Phi^2\Phi^2\Phi^3
-q\tq22\tq11\tq33\Phi^2\Phi^1\Phi^3-q\tq22\tq12\tq33\Phi^2\Phi^2\Phi^3\\
&=abc\begin{pmatrix} 1 & 0&0\\ 0&1&0\\0&0&1\end{pmatrix} \Phi^1\Phi^2\Phi^3
+dbc\begin{pmatrix} 0 & 0&0\\ q&0&0\\0&0&0\end{pmatrix}  \Phi^2\Phi^2\Phi^3\\
&-q abc\begin{pmatrix} 1 & 0&0\\ 0&1&0\\0&0&1\end{pmatrix}  \Phi^2\Phi^1\Phi^3
-q dbc\begin{pmatrix} 0 & 0&0\\ 1&0&0\\0&0&0\end{pmatrix} \Phi^2\Phi^2\Phi^3\\
=&\left(\Phi^1\Phi^2\Phi^3-q  \Phi^2\Phi^1\Phi^3\right) 1\!\!1
\end{split}
\ee
This is proportional to the unit element of the auxiliary space and the superpotential is thus 
invariant.\footnote{Recall that all
the expressions above are in the gauge theory trace, and cyclic permutations of the $\Phi^i$ 
work as usual, since the quantum algebra structure is compatible with the non-abelian structure. E.g. (for $h=0$): 
$\Tr \Phi^1\Phi^2\Phi^3=\Phi^{1a}\Phi^{2b}\Phi^{3c}\Tr(T^aT^bT^c)=
q \Phi^{2b}\Phi^{1a}\Phi^{3c}\Tr(T^aT^bT^c)=q\Phi^{2b}q^{-1}\Phi^{3c}\Phi^{1a}\Tr(T^cT^aT^b)=
\Tr \Phi^2\Phi^3\Phi^1$.}

Of course we did not really have to use this particular matrix representation, we could simply have applied
the full commutation relations of the $\tq ij$ together with those of the $\Phi^i$ to show invariance. 
Thus the above result holds for the general commutation relations in Table \ref{algebrarelations}. In particular,
even though in the general case the matrix representations of $t^i_j$ in some auxiliary space would be
 expected to be infinite-dimensional,  
they are still well-defined as operators and the calculation would go through in a similar way. 


\bibliography{localrefs}
\bibliographystyle{JHEP}

\end{document}